\documentclass[a4paper,12pt]{article}
\usepackage{amsfonts,amsthm,amsmath,amssymb,graphicx}
\numberwithin{equation}{section}

\begin{document}

\newtheorem{definition}{Definition}[section]
\newcommand{\be}{\begin{equation}}
\newcommand{\ee}{\end{equation}}
\newcommand{\bea}{\begin{eqnarray}}
\newcommand{\eea}{\end{eqnarray}}
\newcommand{\LE}{\left[}
\newcommand{\R}{\right]}
\newcommand{\nn}{\nonumber}
\newcommand{\Tr}{\text{Tr}}
\newcommand{\N}{\mathcal{N}}
\newcommand{\G}{\Gamma}
\newcommand{\vf}{\varphi}
\newcommand{\LL}{\mathcal{L}}
\newcommand{\Op}{\mathcal{O}}
\newcommand{\HH}{\mathcal{H}}
\newcommand{\arctanh}{\text{arctanh}}
\newcommand{\up}{\uparrow}
\newcommand{\down}{\downarrow}
\newcommand{\ket}[1]{\left| #1 \right>}
\newcommand{\bra}[1]{\left< #1 \right|}
\newcommand{\ketbra}[1]{\left|#1\right>\left<#1\right|}
\newcommand{\rd}{\partial}
\newcommand{\de}{\partial}
\newcommand{\ba}{\begin{eqnarray}}
\newcommand{\ea}{\end{eqnarray}}
\newcommand{\db}{\bar{\partial}}
\newcommand{\we}{\wedge}
\newcommand{\ca}{\mathcal}
\newcommand{\lr}{\leftrightarrow}
\newcommand{\f}{\frac}
\newcommand{\s}{\sqrt}
\newcommand{\vp}{\varphi}
\newcommand{\hvp}{\hat{\varphi}}
\newcommand{\tvp}{\tilde{\varphi}}
\newcommand{\tp}{\tilde{\phi}}
\newcommand{\ti}{\tilde}
\newcommand{\ap}{\alpha}
\newcommand{\pr}{\propto}
\newcommand{\mb}{\mathbf}
\newcommand{\ddd}{\cdot\cdot\cdot}
\newcommand{\no}{\nonumber \\}
\newcommand{\la}{\langle}
\newcommand{\lb}{\rangle}
\newcommand{\ep}{\epsilon}
 \def\we{\wedge}
 \def\lr{\leftrightarrow}
 \def\f {\frac}
 \def\ti{\tilde}
 \def\ap{\alpha}
 \def\pr{\propto}
 \def\mb{\mathbf}
 \def\ddd{\cdot\cdot\cdot}
 \def\no{\nonumber \\}
 \def\la{\langle}
 \def\lb{\rangle}
 \def\ep{\epsilon}

\begin{titlepage}
\thispagestyle{empty}

\begin{flushright}
NORDITA-2015-73\\
WITS-MITP-016
\end{flushright}

\vspace{.5cm}
\begin{center}
\noindent{\Large \textbf{Entanglement constant for conformal families}}
\vspace{1.5cm}

Pawe{\l} Caputa$^{a}$ and Alvaro Veliz-Osorio$^{b}$

\vspace{0.5cm}
{\it
$^{a}$Nordita, KTH Royal Institute of Technology and Stockholm University,\\
Roslagstullsbacken 23, SE-106 91 Stockholm, Sweden\\
\vspace{0.1cm}
$^{b}$ Mandelstam Institute for Theoretical Physics, School of Physics\\ University of the Witwatersrand, Johannesburg, WITS 2050, South Africa
}

\vskip 2em
\end{center}

\vspace{.5cm}
\begin{abstract}
We show that in 1+1 dimensional conformal field theories, exciting a state with a local operator increases the R\'enyi entanglement entropies by a constant which is the same for every member of the conformal family. Hence, it is an intrinsic parameter that characterises local operators from the perspective of quantum entanglement.
In rational conformal field theories this constant corresponds to the logarithm of the quantum dimension of the primary operator.  We provide several detailed examples for the second R\'enyi entropies and a general derivation.

\end{abstract}

\end{titlepage}

\tableofcontents

%%%%%%%%%%%%%%%%%%%%%%%%%%
\section{Introduction}
%%%%%%%%%%%%%%%%%%%%%%%%%%

Conformal field theories (CFT) in two dimensions play a very important role in the understanding of quantum entanglement (see \cite{CC} for review). For example, in the vacuum state, one can compute entanglement measures like R\'enyi entropies analytically for an arbitrary interval. They are universal and determined by the central charge c and in particular the von-Neumann entropy has the famous $\frac{c}{3}\log L$ scaling with the size of the interval $L$. This relation has been tested quite extensively and serves as an efficient way to numerically determine the central charge for the CFT that governs particular critical points.

\medskip
The next natural step is to explore the entanglement in excited states. A particularly useful protocol to study that, is provided by global or local quenches \cite{cag} where one can follow the time evolution of entanglement measures in a state that differs from an eigenstate globally or locally respectively. Under such evolution, entanglement grows in the system and it is hard to study numerically, nevertheless, the power of (boundary) CFT allows to extract the universal features like the speed of the growth of entanglement. 

\medskip
Recently, inspired by the quench setup, another class of excited states by local operators was proposed in \cite{Nozaki:2014hna}\footnote{see also \cite{Alcaraz:2011tn,Palmai:2014jqa} for a different setup with local operators}. They can be thought of as a milder version of a local quench and the growth of entanglement is less drastic. In fact, at late times, the R\'enyi entanglement entropies saturate to constants that can be used to characterize local operators by the way they change quantum entanglement in a given state. Detailed reviews can be found in \cite{Caputa:2014vaa,Nozaki:2014uaa} and further results and applications to finite temperature and holography in \cite{Shiba:2014uia,He:2014mwa,Guo:2015uwa,Caputa:2014eta,Asplund:2014coa,Caputa:2015waa}. 

\medskip
For primary operators in 1+1 dimensional rational CFTs, the growth of the entanglement of an interval can be computed analytically and it was proved to be the logarithm of the quantum dimension of the local operator \cite{He:2014mwa}. Since in this setup we can uniquely decompose an operator into left and right movers (chiral, anti-chiral parts) this constant is equivalent the the left-right entanglement in this excited state. Intuitively, this is explained by the quasi-particle picture for the propagation of entanglement where the excitation can be thought of as  insertion of the EPR like pair. Quasiparticles propagate in opposite directions and once one of them is in the entangling region it increases the entanglement with the rest by the entanglement of the pair. This picture is supported by the one point functions of the energy in locally excited state \cite{NNT}.

\medskip
In this work we also focus on states excited by local operators in 1+1 d CFTs with Virasoro symmetry, and in the search for universal features of entanglement we ask how the above picture changes for descendant operators. As we will show, one-point functions of the energy behave differently in the states excited by descendants. Moreover, their conformal transformations, as well as their correlators, are much more complicated than for primaries and the appearance of the quantum dimension is not obvious. Nevertheless, we will show for descendants up to level 2 and provide a general  argument that the constant contribution to the R\'enyi entanglement entropies is the same for the entire conformal family.

\medskip
This paper is organized as follows. In section 2 we briefly review the replica method with local operators in 2d CFT \cite{He:2014mwa} and  provide an example that illustrates the CFT technology. In section 3 we compute one-point functions of the energy density and the increase in the second Renyi entropy for descendant operators up to level 2. In section 4 we outline a general argument for why the contribution to the R\'enyi entanglement is a characteristic of the entire conformal family. In section 5 conclude and discuss the Schmidt decomposition and possible connection to topological entanglement entropy. Several details of the analysis are moved to the appendices.

%%%%%%%%%%%%%%%%%%%%%%%%%%
\section{Replica trick for locally excited states}\label{sec:replica}
%%%%%%%%%%%%%%%%%%%%%%%%%%

The entanglement entropy of a subsystem quantifies the amount of information that we would forfeit if 
we were to loose access to the rest of the system. Imagine that the system of interest is in a pure state $|\Psi\rangle$ and suppose that we wish to quantify the entanglement between a subsystem $A$ and its complement $\bar A$.
As a first step we find the reduced density matrix obtained by tracing out the degrees of freedom in $\bar A$
\be
\rho_A=\Tr_{{\cal H}_{\bar A}}|\Psi\rangle\langle \Psi|\, .
\ee
Once we have constructed this matrix, we notice that if there is any entanglement between the
degrees of freedom in $A$ and those in $\bar A$ the system appears, to an observer having access only to $A$, to be in a mixed state.
If that is the case, then the Von Neumann entropy of $\rho_A$
\be
S_A=-\Tr(\rho_A\log \rho_A)\, ,\label{eq:def EE}
\ee
is non-vanishing. We refer to this quantity as the entanglement entropy of $A$. In practice, we use the so-called the replica trick \cite{CC} and compute instead the R\'enyi entropies
\be
S_A^{(n)} =\frac{1}{1-n}\log\Tr\left[\rho^n_A\right]\,.
\ee
The entanglement entropy \eqref{eq:def EE} can be extracted from the above expression by taking the $n\rightarrow 1$ limit. These R\'enyi entropies provide interesting measures of entanglement on their own right e.g. the \textit{min entropy} $S_A^{(\infty)}$ and the \textit{purity} $S_A^{(2)}$ \cite{NCH}.

\medskip
It is natural to wonder what would be the effect on the entanglement between different parts of a system if we were to perturb it in some way. In this work we focus on the case where these perturbation is due to the insertion of a local operator. Recently, the replica method to compute the R\'enyi  entropies has been generalized to deal with these scenarios. Below we summarize the relevant formulas and refer the reader for more details to \cite{Nozaki:2014hna,Caputa:2014vaa,He:2014mwa}.

\medskip
Consider the ground  state of a 1+1 dimensional CFT on a line. We split the space into a finite interval $A=[l_1,l_2]$ of length $L\equiv l_2-l_1$ and its complement. Then at $t=0$ we insert a local operator ${\cal O}$ at $x=-l$ and let the system evolve. Without losing generality, we can shift the interval to $A=[0,L]$. 
The resulting density matrix is given by
\bea
\rho(t, l, L,\epsilon)&=&{\mathcal N}\cdot e^{-iHt}e^{-\ep H}{\cal O}(0,-l)|0\lb\la 0|{{\cal O}}^{\dagger}(0,-l)e^{-\ep H}e^{iHt} \nn\\
&\equiv& {\mathcal N}\cdot {\cal O}(w_2,\bar{w}_2)|0\lb\la 0|
{{\cal O}}^{\dagger}(w_1,\bar{w}_1)\,.  \label{denmat}
\eea
where $\epsilon\ll 1$ plays the role of an UV regulator\footnote{Analogously to the local quench setup, we assume that all length and time scales are much larger
than $\epsilon$.} and ${\cal N}$ is a normalization that ensures that $\Tr\,\rho=1$. In the second line the \textit{insertion points} are defined as
\ba
&& w_1=i(\epsilon -it)-l, \ \ w_2 = -i(\epsilon+it)-l, \nn\\
&& \bar{w}_1=-i(\ep-it)-l,\ \ \bar{w}_2=i(\epsilon+it)-l.\label{points}
\ea
Hereafter, we omit the dependence on $ l$, $ L$, $\epsilon$ and write simply $\rho(t)$. 

\medskip
The increase in the $n$-th  R\'enyi entanglement entropy of the interval $A$ wrt to the ground state due to the local operator is given by \cite{Nozaki:2014hna}
\bea
\Delta S_A^{(n)}
\equiv\frac{1}{1-n}\log\left( \frac{\langle {\cal O}(w_1, \bar{w}_1){\cal O}^{\dagger}(w_2, \bar{w}_2)\cdots {\cal O}(w_{2n-1}, \bar{w}_{2n-1}){\cal O}^{\dagger}(w_{2n}, \bar{w}_{2n})\rangle_{\Sigma_n}}{\left(\langle {\cal O}^{\dagger}(w_1, \bar{w}_1){\cal O}(w_2, \bar{w}_2)\rangle_{\Sigma_1}\right)^n}\right),\label{eq:Delta}
\eea
where $\Sigma_n$ is the $n$-sheeted surface with cuts on each copy corresponding to $A$, and $\Sigma_1$ is a single copy with an interval cut $A$. Thus, we are faced with the task of evaluating a $2n$-point function on $\Sigma_n$. As shown in \cite{He:2014mwa}, this can be computed by using the uniformization map 
\be
z^n=\frac{w}{w-L}\label{map}\,,
\ee
that takes $\Sigma_n$ to the complex plane. Notice that the transformation properties of ${\cal O}$ under the above map are expected to play an important role.

\medskip
 Let us focus on the change in the second R\'enyi entropy $\Delta S^{(2)}_A$. In this case we are expected to evaluate the operator's four-point function on a two-sheeted surface. We use \eqref{map} with $n=2$ to map $\Sigma_2$ to the plane, where conformal symmetry ensures that these correlator can be written in terms of the cross-ratios of the four insertion points. Notice that under the uniformization map the insertion points \eqref{points} become $z_4=-z_2$ and $z_3=-z_1$. 
From equations \eqref{points} and \eqref{map} and in the limit $\epsilon\to 0$ the cross ratios are given by \cite{He:2014mwa}
\be
z=\frac{z_{12}z_{34}}{z_{13}z_{24}}\simeq 1-\f{L^2\ep^2}{4(l-t)^2(L+l-t)^2},\qquad  \bar{z}\simeq \f{L^2\ep^2}{4(l+t)^2(L+l+t)^2}~,\label{lik}
\ee
provided that $t\in[l,L+l]$. 
On the other hand, if $t\notin[l,L+l]$, then the cross-ratios read
\be
z\simeq\f{L^2\ep^2}{4(l-t)^2(L+l-t)^2},\qquad \bar{z}=\frac{\bar{z}_{12}\bar{z}_{34}}{\bar{z}_{13}\bar{z}_{24}}\simeq\f{L^2\ep^2}{4(l+t)^2(L+l+t)^2},
\label{etim}
\ee
Hence, there are two possibilities, $(z,\bar{z})\simeq(1,0)$ for $t\in[l,L+l]$, while $(z,\bar{z})\simeq(0,0)$ otherwise.

\medskip
The authors of \cite{He:2014mwa} showed that if ${\cal O}$ is a primary operator, then
\be
\Delta S^{(2)}_A(z,\bar z)=-\log\left[|z(1-z)|^{4h}{\cal G}(z,\bar{z})\right]\,.
\ee
where ${\cal G}(z,\bar{z})$ is the canonical four point function of the operators on the plane. 
As discussed above, we must evaluate this result for the regimes $(z,\bar z)\simeq(0,0)$ and $(z,\bar z)\simeq(1,0)$. In the first regime there is no increase in entanglement
\be
\Delta S^{(2)}_A(0,0)=0.
\ee
Now, since the two regimes are related to one another by the modular transformation $z\rightarrow 1-z$, in rational CFT, it follows that 
\be
\Delta S^{(2)}_A(1,0)=-\log\,F_{00}[{\cal O}]\,,\label{DS2r}
\ee
where $F_{00}[{\cal O}]$ is the component of fusion matrix for the identity \cite{DiF}. Moreover,  $F_{00}[{\cal O}]$  corresponds to the inverse 
of the \textit{quantum dimension} $d_{\cal O}$ of the operator \cite{Moore:1988ss}. Therefore, one gets
\be
\Delta S^{(2)}_A=\left\{
	\begin{array}{ll}
		0 & \mbox{if }t\notin[l,L+l]\\
		\log d_{\cal O} & \mbox{if }t\in[l,L+l]
	\end{array}\label{eq:primary}
\right.
\ee
This behavior of the R\'enyi entropies has been interpreted in terms of a quasi-particle picture where the insertion of the local operator corresponds to creation of the EPR-like pair whose members propagate in the opposite directions (to the left and right). A non-trivial contribution to the R\'enyi entropy comes from times when one of the members is inside the interval $A$ while the other one is outside.

\medskip
In this work we will test these results and quasi-particle picture further by considering excitations by descendant operators. 

%%%%%%%%%%%%%%%%%%%%%%%%%%
%%%%%%%%%%%%%%%%%%%%%%%%%%
\subsection{Example: EPR-primary}\label{sec:EPRprimary}
%%%%%%%%%%%%%%%%%%%%%%%%%%
%%%%%%%%%%%%%%%%%%%%%%%%%%

In this section we consider a simple illustrative example of a primary operator that is a generalization of the operators studied in \cite{He:2014mwa}, which will also serve to check our results for descendants.
Let us start with a state locally excited by the primary operator
\be
{\cal O}_{\alpha}=\sqrt{a}\,e^{i\sqrt{2}\alpha\phi}\pm\sqrt{1-a}\,e^{-i\sqrt{2}\alpha\phi}\, ,\label{eq:EPR primary}
\ee
where $\phi$ is a free massless scalar and $a\in[0,1]$. It is easy to verify that the conformal dimension of the operator is $h=\bar{h}=\alpha^2$ . Hereafter, we refer to the operator \eqref{eq:EPR primary} as the \textit{EPR-primary}.

%%%%%%%%%%%%%%%%%%%%%%%%%%
\subsubsection*{Replica method}
%%%%%%%%%%%%%%%%%%%%%%%%%%

For the sake of simplicity we consider the second R\'enyi entropy. As discussed in the previous section, since ${\cal O}_{\alpha}$  is a primary operator the increase of $ S^{(2)}_A$  is given by
\be
\Delta S^{(2)}_A=-\log\left[|z(1-z)|^{4h}{\cal G}_{\alpha}(z,\bar{z})\right]\,,
\ee
 where ${\cal G}_{\alpha}(z,\bar{z})$ is defined by the four-point correlator of the EPR operators on the complex plane via
 \be
\langle {\cal O}_{\alpha}(z_1,\bar{z}_1){\cal O}^\dagger_{\alpha}(z_2,\bar{z}_2){\cal O}_{\alpha}(z_3,\bar{z}_3){\cal O}^\dagger_{\alpha}(z_4,\bar{z}_4)\rangle=|z_{13}z_{24}|^{-4h}{\cal G}_{\alpha}(z,\bar{z})\,,\label{eq:fpf plane}
\ee
or more explicitly we define
\be
{\cal G}_{\alpha}(z,\bar{z})= \langle {\cal O}_\alpha|{\cal O}_{\alpha}(z,\bar{z}){\cal O}_{\alpha}(1,1)|{\cal O}_{\alpha}\rangle\equiv|z(1-z)|^{-4h}\mathcal{F}_{\alpha}(z,\bar{z})\,.
\ee
The function $\mathcal{F}_{\alpha}(z,\bar{z})$, which encodes the non-singular part of  $\mathcal{G}_{\alpha}(z,\bar{z})$ can be computed explicitly for the EPR operators (see Appendix \ref{App:CO}) and it is given by
\be
\mathcal{F}_{\alpha}(z,\bar{z})=a^2+(1-a)^2+2a(1-a)\left(|z|^{8h}+|1-z|^{8h}\right).
\ee
In terms of this function, the change in the R\'enyi entropy takes the compact form 
\be
\exp(-\Delta S^{(2)}_A)=\mathcal{F}_{\alpha}(z,\bar{z}).
\ee

\medskip
Finally, for times $t\notin [l,L+l]$ we have both $(z,\bar{z})\simeq (0,0)$ and the change in the second R\'enyi entropy vanishes. On the other hand for $t \in [l,L+l]$ the cross-ratios are approximately $(z,\bar{z})\simeq (1,0)$ , which gives a non-trivial contribution to the second R\'enyi entropy
\be
\Delta S^{(2)}_A\simeq \left\{
\begin{array}{ll}
0&\qquad t\notin [l, L+l] \\
-\log\left[a^2+(1-a)^2\right] &\qquad t\in [l, L+l] \label{EPR entropy}
\end{array}
\right.
\ee
Notice that the non-vanishing contribution to the second R\'enyi is maximized by an EPR primary with  $a=\frac{1}{2}$. 

%%%%%%%%%%%%%%%%%%%%%%%%%%%
\subsubsection*{Left-right entanglement}
%%%%%%%%%%%%%%%%%%%%%%%%%%

In the following, we show how the same result can be obtained from the entanglement between left and right moving sectors in the state
\be
{\cal O}_{\alpha}\ket{0}_L\otimes\ket{0}_R=\sqrt{a}\ket{e^{i\sqrt{2}\alpha\phi_L}}_L\otimes\ket{e^{i\sqrt{2}\alpha\phi_R}}_R\pm\sqrt{1-a}\ket{e^{-i\sqrt{2}\alpha\phi_L}}_L\otimes\ket{e^{-i\sqrt{2}\alpha\phi_R}}_R\, ,\label{STATELR}
\ee 
where we used the decomposition of the scalar field $\phi(t,x)=\phi_L(t+x)+\phi_R(t-x)$.
Tracing out the right-moving sector we find that the reduced density matrix of the left-movers, which is simply
\be
\rho_L=\text{diag}\{a,1-a\}\,.\label{DMEPR}
\ee
 Hence, the n-th R\'enyi entropy is given by
\be
S^{(n)}_{L}=\frac{1}{1-n}\log \Tr \rho^n_L=\frac{1}{1-n}\log\left[a^n+(1-a)^n\right],\label{RenLR}
\ee
which for $n=2$ reproduces Eq. \eqref{EPR entropy}.
\begin{figure}[h!]	
  \centering
  \includegraphics[width=11cm]{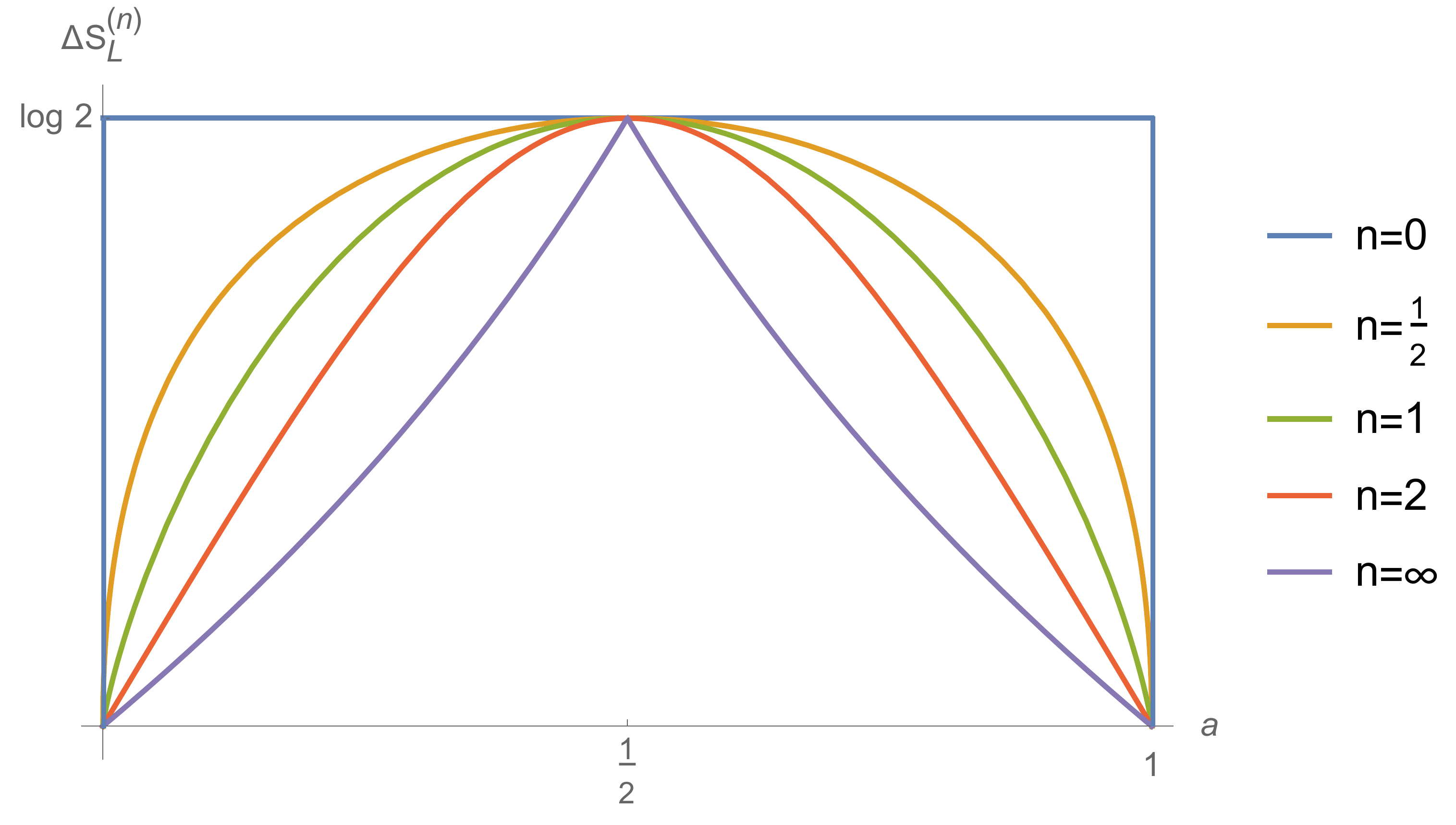}
  \caption{Finite value of the R\'enyi entropies as function of $a$ for different replica numbers $n$ in a state excited by operator \eqref{eq:EPR primary} }\label{Renyis}
\end{figure}

\medskip
Clearly the left-right decomposition of the excited state gives a quick way to find  the reduced density matrix. Once we have it\footnote{Generically the left-right decomposition of the operator is not as simple as in our example and depends on the structure of the Hilbert space of the CFT}, we can study the R\'enyi entropies for different values of $n$ and see how they are affected by the local excitation \eqref{eq:EPR primary}. For instance the Hartley entropy ($n=0$) is given by $S^{(0)}_{L}=\log(2)$  and sets an $a$-independent upper bound for all the R\'enyi entropies. It corresponds to the log of the dimension of the reduced density matrix $\rho_L$. In turn, the entanglement entropy ($n=1$) equals the binary entropy
\be
S^{(1)}_L=H_2(a)\equiv-a\log(a)-(1-a)\log(1-a).
\ee
Finally the min entropy ($n\to\infty$) is given by the inverse of the logarithm of the largest eigenvalue of $\rho_L$
\be
S^{(\infty)}_L=-\log\max\left[a,1-a\right].
\ee
All the R\'enyi entropies for $n\geq 0$ are sensitive to the special points $a=0$ and $a=1$ for which they vanish, reflecting the fact that for these values \eqref{STATELR} becomes a product state. On the other hand, for $a=\frac{1}{2}$ they saturate the inequality set by the Hartley entropy as expected from 
this maximally entangled mixture (see Fig \ref{Renyis} for comparisons).

\medskip
Let us stress that for general operators like EPR with different $a$ each R\'enyi entropy gives a different constant dependent on $a$ and only for $a=1/2$ they coincide. This is the case studied in \cite{He:2014mwa} and corresponds to the sigma operator in the Ising model. Hence, we expect that only for the rational CFTs all the R\'enyi entanglement entropies will increase by the same constant.

%%%%%%%%%%%%%%%%%%%%%%%%%%
\subsubsection*{Energy density}
%%%%%%%%%%%%%%%%%%%%%%%%%%

 The result \eqref{EPR entropy} has been interpreted as the effect on entanglement due to an EPR pair whose members are receding from each other at the speed of light. To test this intuition, one could compute the time evolution for the expectation value of the energy density 
\be
T_{tt}(x,\bar{x})=-(T(x)+\bar{T}(\bar{x}))\,,
\ee
in the excited state corresponding to ${\cal O}_\alpha$ \cite{Caputa:2014vaa}. 
 The three-point function of the primary operators with the stress tensor is universal and the final answer is given by 
\bea
\langle T_{tt}(x,x)\rangle_{{\cal O}_\alpha}&\equiv&\frac{\langle {\cal O}^{\dagger}_{\alpha}(w_2,\bar{w}_2)T_{tt}(x,x){\cal O}_{\alpha}(w_1,\bar{w}_1)\rangle}{\langle {\cal O}_{\alpha}(w_1,\bar{w}_1){\cal O}^{\dagger}_{\alpha}(w_2,\bar{w}_2)\rangle}\nn\\
&=&\frac{4h\,\epsilon^2}{((x+l-t)^2+\epsilon^2)^2}+\frac{4\bar{h}\,\epsilon^2}{((x+l+t)^2+\epsilon^2)^2}.
\eea
Indeed, the above expression describes two wave packets of width $\epsilon$ propagating in the opposite directions from the insertion point. The total energy injected to the system with the operator is equal to $E=\frac{2\pi(h+\bar{h})}{\epsilon}$.

\begin{figure}[h!]	
  \centering
  \includegraphics[width=11cm]{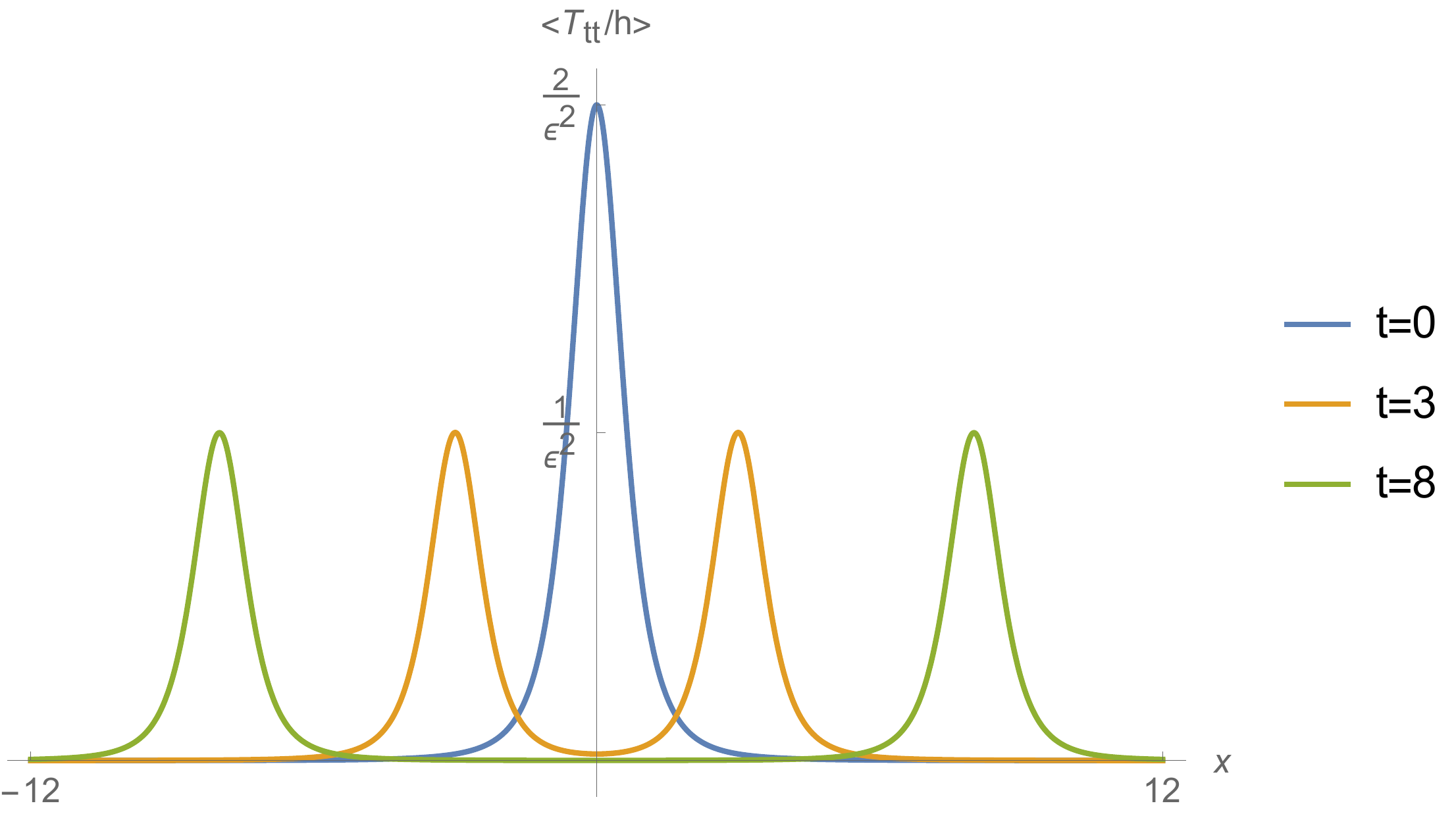}
  \caption{Evolution of the energy density for three different times. Plot for $l=0$ and $\epsilon=1$.}\label{EnPr}
\end{figure}

\medskip
The evolution of the energy density profile strongly supports the EPR interpretation of the operator excitation since the two wave packets resemble the pair (see Fig. \ref{EnPr}). Notice that the above computation is very robust and insensitive to the details of the primary operator, for example, the value of $a$ in \eqref{eq:EPR primary}.

%%%%%%%%%%%%%%%%%%%%%%%%%%
%%%%%%%%%%%%%%%%%%%%%%%%%%
\section{Descendants examples}
%%%%%%%%%%%%%%%%%%%%%%%%%%
%%%%%%%%%%%%%%%%%%%%%%%%%%i

In this section we compute the increase in the second R\'enyi entropy for few simple descendants.  As we shall see, although some modifications on the energy density arise, the behavior of the R\'enyi entropy remains the same.

%%%%%%%%%%%%%%%%%%%%%%%%%%
\subsection{Energy-momentum tensor}
%%%%%%%%%%%%%%%%%%%%%%%%%%

Arguably, the most important non-primary operator in a CFT is the stress-energy tensor $T(z)$. Recall that $T(z)$ is a descendant of the identity operator with weight $h=2$. It must be clear that the insertion of the identity operator into the system doesn't cause any  modification in the R\'enyi entropy. It is with the stress-energy tensor that we start our study of the effect of the insertion of descendants on entanglement. 
In the following, we compute the change of the second R\'enyi entropy $\Delta S^{(2)}$ due to the insertion of the chiral part of the stress-energy tensor $T(z)$.  For this case, equation \eqref{eq:Delta} reads 
\be
\exp\left(-\Delta S_A^{(2)}(z)\right)=\frac{\langle T(w_1) T(w_2)T(w_3) T(w_4)\rangle_{\Sigma_2}}{\left(\langle T(w_1)T(w_2)\rangle_{\Sigma_1}\right)^2}.\label{eq:Delta T}
\ee
In order to compute the numerator of the above equation, we map the two-sheeted surface to the plane using the map \eqref{map}, under which $T(z)$ transforms as 
\be
 T(w_i)=\frac{\left(1-z_i^2\right)^2}{4\, z_i^2L^2}\left[ T(z_i)+\frac{c}{8z_i^2}\right]\equiv \alpha_i\left[ T(z_i)+\beta_i\right].
\ee
On the plane, the four-point function of the energy-momentum tensor is given by
\begin{align}
\langle T(z_1)T(z_2)T(z_3)T(z_4)\rangle= \left(z_{13}z_{24}\right)^{-4} {\cal G}_T(z)\equiv\left(z_{13}z_{24}z(1-z)\right)^{-4}\mathcal{F}_T(z),\label{4ptT}
\end{align}
where \cite{Osborn:2012vt}
\be
{\cal G}_T(z)=\frac{c^2}{4}\left(1+\frac{1}{z^4}+\frac{1}{(1-z)^4}\right)+2c\frac{(1-z(1-z))}{z^2(1-z)^2}\label{cblock T}, 
\ee
and
\begin{align}
\mathcal{F}_T(z)&=\frac{c^2}{4}\left[z^4(1-z)^4+(1-z)^4+z^4\right]+2c\, z^2(1-z)^2\left[1-z(1-z)\right]\,,\label{eq:FT}
\end{align}
where the latter is constructed from ${\cal G}_T(z)$ by extracting the leading singular behavior.

\medskip
The contributions to \eqref{eq:Delta T} coming from the Jacobian and the denominator can be written in terms of the cross ratios and read
\be
\frac{\prod^4_i\alpha_i}{z^4_{13}z^4_{24}\langle T(w_1)T(w_2)\rangle^2_{\Sigma_1}}=\frac{z^4(1-z)^4}{c^2/4}.
\ee
Hence, we find that
\be
\exp\left(-\Delta S_A^{(2)}(z)\right)=\frac{z^4(1-z)^4}{c^2/4}\left[{\cal G}_T(z)+\frac{c^4}{16}+\frac{c^3}{4}\frac{\left[1 - z( 1-z) \right]^2}{z^2(1-z)^2}-\frac{2c^2}{z(1-z)}\right]\,,\label{Tres}
\ee
where the terms in the bracket originate from the non-vanishing correlators of products of $T(z_i)+\beta_i$.

\medskip
Once again, as we take $\epsilon \to 0$, we either have $z\simeq 0$ for $t\notin [l,L+l]$ or $z\simeq1$ for $t\in [l,L+l]$.  
Notice, that the pre-factor of \eqref{Tres} cancels the divergence of ${\cal G}_T(z)$ at those points. This is, moreover, the highest degree divergence inside the brackets. To make this fact manifest, we rewrite \eqref{Tres} as
\be
\exp\left(-\Delta S_A^{(2)}(z)\right)=\frac{4}{c^2}\left[{\cal F}_T(z)+z^4(1-z)^4\left(\frac{c^4}{16}+\frac{c^3}{4}\frac{\left[1 - z( 1-z) \right]^2}{z^2(1-z)^2}-\frac{2c^2}{z(1-z)}\right)\right]\,.
\ee
Therefore, we find 
\be
\exp\left(-\Delta S_A^{(2)}\right)\simeq \left\{
\begin{array}{ll}
\frac{4}{c^2}{\cal F}_T(0)&\qquad t\notin [l,l+L]\nonumber \\
\frac{4}{c^2}{\cal F}_T(1) &\qquad t\in [l,l+L] \label{2Rex}
\end{array}
\right.
\ee
From equation \eqref{eq:FT} we see that 
\be
{\cal F}(z\to 0)\,=\,{\cal F}(z\to1)\,=\,\frac{c^2}{4},
\ee
thus
\be
\Delta S_A^{(2)}=0
\ee
for all times. This way we confirm that both, primary $\mathbb I$ and descendant $T(z)$ lead to the same behavior, $\Delta S_A^{(2)}=0$, for any 1+1d CFT (not necessarily rational). As we will show, this phenomenon is not a coincidence and we will provide further evidence for it in the remainder of this work. 

\medskip
A clarification at this point is in order. From the above formulas we can notice an order of limits conflict if we consider CFTs with large central charge. Namely in order to extract the interesting constant (in this case equal to zero) we first have to take $\epsilon\to 0$ and then $c\to\infty$ (for the opposite interesting order see App \ref{App:Lc}).

%%%%%%%%%%%%%%%%%%%%%%%%%%
\subsection{Derivative of a primary}
%%%%%%%%%%%%%%%%%%%%%%%%%%

In this section we consider a state locally excited by the first descendant of a primary operator
${\cal O}$. We denote the descendant operator by
\be
\partial{\cal O}(z,\bar{z})=L_{-1}{\cal O}(z,\bar{z})= \partial_z {\cal O}(z,\bar{z})\,.\label{Desa}
\ee
Clearly, the conformal dimensions of $\partial{\cal O}$ are $(h+1,\bar{h})$, where $(h,\bar h)$ are those corresponding to the primary. Let us start by computing the evolution of the energy density  in states locally excited by \eqref{Desa}. As mentioned above, the expectation value of the energy density is universal. As a matter of fact, we will only need the OPE between the stress tensor and the derivative of a primary field which reads \cite{DiF}
\bea
T(x)\partial{\cal O}(z_i,\bar{z}_i)&\sim& \frac{2h {\cal O}(z_i,\bar{z}_i)}{(x-z_i)^3}+\frac{(h+1)\partial_{z_i}{\cal O}(z_i,\bar{z}_i)}{(x-z_i)^2}+\frac{\partial^2_{z_i}{\cal O}(z_i,\bar{z}_i)}{x-z_i}\,,\label{OPEpd}
\eea
whereas the OPE with $\bar{T}(\bar{x})$ is the same as for the primary.

\medskip

\begin{figure}[h!]	
  \centering
  \includegraphics[width=11cm]{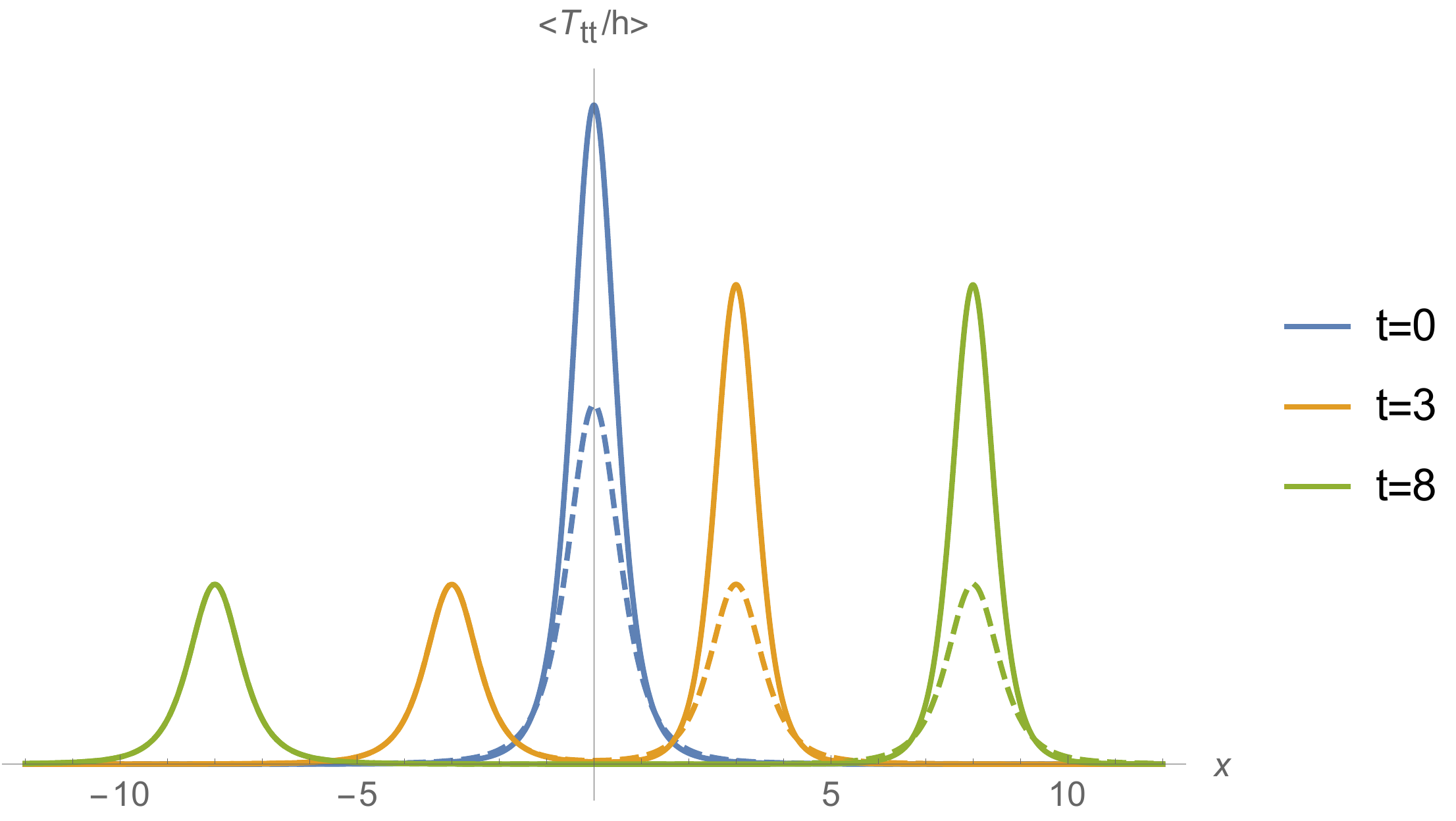}
  \caption{Evolution of the energy density for three different times in the state excited by the descendant \eqref{Desa}. For comparison, the dashed lines show the evolution in the state excited by a primary. Plot for $l=0$ and $\epsilon=1$.}\label{EnergyEx}
\end{figure}

The correlators with the stress tensor and descendants can be effectively written in terms a differential operator acting on the correlation functions of the primary operators (see also section \ref{DesL2}). More precisely, we can write the three-point correlator as a sum over the residues from the OPEs of $T$ and $\bar{T}$ with $\partial_i\mathcal{O}_i$, $i=1,2$, taken inside the correlator and then pull the derivatives in front, such that the operator acts on the two-point function of the primaries $\mathcal{O}$. Applying the differential operator and inserting the points \eqref{points}, the expectation value of the energy density can be written as
\be
\langle T_{tt}(x,x)\rangle_{\partial O}=\langle T_{tt}(x,x)\rangle_{O}+\frac{4\epsilon^2\left[(1-4h)(x+l-t)^2+(1+4h)\epsilon^2\right]}{(2h+1)\left[(x+l-t)^2+\epsilon^2\right]^3},
\ee
and the total energy injected into the system equals
\be
E_{\partial O}=\int dx\, \langle T_{tt}(x,x)\rangle_{\partial O}=\frac{2\pi(h+1+\bar{h})}{\epsilon}\,,
\ee
as expected. Once again, the evolution of the energy density corresponds to that of two wave packets propagating in opposite directions from the insertion point. However, in the present case there is an asymmetry in the amount of energy propagating on each side.  Figure \ref{EnergyEx} clearly displays the asymmetry, where the right moving wave packet is seen to carry more energy. Naively one could expect that this different behavior of the energy density will imply the different constant increase in the R\'enyi entanglement entropies. We will show below that this turns out not to be correct.

\medskip
Let us first confront this with the left-right entanglement of the excited state. Formally, we can also think of the descendant operator \eqref{Desa} as acting on the left and right vacuum so that we have a new entangled state\footnote{For simplicity we suppress the $L/R$ indices that indicate left or right movers in the Minkowski signature or the functions of $z$ or $\bar{z}$ in Euclidean.}
\be
\partial_zO_{\alpha}\ket{0}\otimes\ket{\bar{0}}=i\sqrt{2}\alpha\left[\sqrt{a}\ket{\partial\phi e^{i\sqrt{2}\alpha\phi}}\otimes\ket{e^{i\sqrt{2}\alpha\bar{\phi}}}-\sqrt{1-a}\ket{\partial\phi e^{-i\sqrt{2}\alpha\phi}}\otimes\ket{e^{-i\sqrt{2}\alpha\bar{\phi}}}\right]
\ee
The normalized density matrix of the holomorphic or anti-holomorphic (left or right) movers for this state becomes \eqref{DMEPR} and therefore the corresponding R\'enyi entropies are again equal to \eqref{RenLR}. This suggests that, despite the different evolution of the energy density, the contribution to the entanglement R\'enyi entropy should be the same as for the primary operator.

\medskip
Now, we proceed to calculate the change in the second R\'enyi entropy wrt to the vacuum due to the insertion of the descendant operator $ \partial{\cal O}$. This can be obtained by computing
\be
\exp\left(-\Delta S_A^{(2)}(z,\bar z)\right)=\frac{\langle\partial{\cal O}(w_1,\bar w_1) \partial{\cal O}(w_2,\bar w_2)\partial{\cal O}(w_3,\bar w_3) \partial{\cal O}(w_4,\bar w_4)\rangle_{\Sigma_2}}{\left(\langle \partial{\cal O}(w_1,\bar w_1) \partial{\cal O}(w_2,\bar w_2)\rangle_{\Sigma_1}\right)^2}\,.\label{delta partial}
\ee
The first step is to map the two-sheeted surface $\Sigma_2$ to the plane using the map \eqref{map}. One must proceed with caution since descendant operators transform differently form primaries under conformal transformations (see App.B). For the first descendant of a primary we have 
\bea
\partial{\cal O}(w_i,\bar{w}_i)&=&(w'_i)^{-(h+1)}(\bar w'_i)^{-\bar h}\left[\partial {\cal O}(z_i,\bar{z}_i)-h\frac{w_i''}{w_i'}{\cal O}(z_i,\bar{z}_i)\right]\nonumber\\
&\equiv&\;\alpha_i\bar\alpha_i\left[ \partial{\cal O}(z_i,\bar{z}_i)+\beta_i\,{\cal O}(z_i,\bar{z}_i)\right].
\eea
Schematically, the correlator in the numerator of Eq. \eqref{delta partial} can be written in terms of correlators on the plane as
\bea
&&\langle\partial{\cal O}(w_1,\bar w_1)...\partial{\cal O}(w_4,\bar w_4)\rangle_{\Sigma_2}=\prod_{i=1}^4 \alpha_i\bar\alpha_i\bigg(\langle {\partial\cal O}\partial{\cal O}\partial{\cal O}\partial{\cal O}\rangle+\sum_j \beta_j \langle {\cal O}\partial{\cal O}\partial{\cal O}\partial{\cal O}\rangle+\nonumber\\
&&\sum_{j,k} \beta_j \beta_k \langle {\cal O}{\cal O}\partial{\cal O}\partial{\cal O}\rangle+ \sum_{j,k,l} \beta_j\beta_k \beta_l \langle {\cal O}{\cal O}{\cal O}\partial{\cal O}\rangle+  \prod_{i=1}^4\beta_i\; \langle {\cal O}{\cal O}{\cal O}{\cal O}\rangle\bigg)\,. 
\eea
 
\medskip
The four-point function of the primary operator ${\cal O}$ is given by Eq. \eqref{eq:fpf plane},
whereas that for the first descendant takes the form\footnote{Notice that this form assumes $z_3=-z_1$ and $z_4=-z_2$ and for general points one cannot write the four point correlator of descendants this way.}
\be
  \langle\partial{\cal O}(z_1,\bar z_1)\partial{\cal O}(z_2,\bar z_2)\partial{\cal O}(z_3,\bar z_3)\partial{\cal O}(z_4,\bar z_4)\rangle=\left(4z_1 z_2\right)^{-2(h+1)}\left(4\bar z_1\bar z_2 \right)^{-2 \bar h} {\cal G}_{\partial\cal O}(z,\bar z)\,.
\ee
The expression for ${\cal G}_{\partial\cal O}$ can be found in appendix \ref{App:CB}, equations  \eqref{G derivative} and \eqref{G derivative coefficients}.
On the other hand, the two point function of $\partial\cal O$ is simply
\be
\langle  \partial {\cal O}(w_1,\bar w_1) \partial {\cal O}(w_2,\bar w_2)\rangle_{\Sigma_1}=-2 h(1 + 2 h)   w_{12}^{-2(h+1)}\bar w_{12}^{-2\bar h}\, .\label{2pf}
\ee
Now, we split the RHS of equation \eqref{delta partial} into two pieces
\begin{align}
\exp\left(-\Delta S_A^{(2)}(z,\bar z)\right)={\cal N}_{\partial{\cal O}}\left({\cal G}_{\partial\cal O}(z,\bar z)+\Xi (z,\bar z)\right)\,,\label{Delta G 1}
\end{align}
where ${\cal N}_{\partial{\cal O}}$ is an overall pre-factor given by
\be
{\cal N}_{\partial{\cal O}}=\frac{ \left(4z_{1}z_{2}\right)^{-2(h+1)}\left(4\bar z_{1}\bar z_{2}\right)^{-2 \bar h}\prod_{i=1}^4 \alpha_i\bar\alpha_i}{\left(\langle\partial {\cal O}(w_1,\bar w_1)\partial {\cal O}(w_2,\bar w_2)\rangle_{\Sigma_1}\right)^2}\,,
\ee
and $\Xi$ contains all the terms of lower order in derivatives, i.e. 
\begin{align}
\Xi (z,\bar z)= &\left(4z_{1}z_{2}\right)^{2(h+1)}\left(4\bar z_{1}\bar z_{2}\right)^{2 \bar h}\bigg(\sum_j \beta_j \langle {\cal O}\partial{\cal O}\partial{\cal O}\partial{\cal O}\rangle+\sum_{j,k} \beta_j \beta_k \langle {\cal O}{\cal O}\partial{\cal O}\partial{\cal O}\rangle\nonumber\\
& \sum_{j,k,l} \beta_j\beta_k \beta_l \langle {\cal O}{\cal O}{\cal O}\partial{\cal O}\rangle+  \prod_{i=1}^4\beta_i\; \langle {\cal O}{\cal O}{\cal O}{\cal O}\rangle\bigg)\,.
\end{align}

\medskip
To find the change of the second R\'enyi entropy,  we use the Jacobian of the uniformization map \eqref{map} and the two point function \eqref{2pf} to show that the pre-factor can be written in terms of the cross ratios as
\begin{equation}
{\cal N}_{\partial{\cal O}}=\frac{((1 - z) z)^{2 (1 + h)} ((1 -\bar z)\bar  z)^{2\bar h}}{4 h^2 (1 + 2 h)^2}.
\end{equation}
Therefore, we have
\begin{align}
\exp\left(-\Delta S_A^{(2)}(z,\bar z)\right)=&\frac{((1 - z) z)^{2 (1 + h)} ((1 -\bar z)\bar  z)^{2\bar h}}{4 h^2 (1 + 2 h)^2}\left[ {\cal G}_{\partial\cal O}(z,\bar z)+\Xi (z,\bar z)\right]\,.\label{Delta G}
\end{align}
For primary operators we have decomposed ${\cal G}_{\cal O}$ into
\be
{\cal G}_{\cal O}(z,\bar z)=(z(1-z))^{-2h}(\bar z(1-\bar z))^{-2\bar h}{\cal F}_{\cal O}(z,\bar z),
\ee
where ${\cal F}_{\cal O}$ is regular. Analogously, for the descendant we define
\be
{\cal G}_{\partial\cal O}(z,\bar z)=(z(1-z))^{-2(h+1)}(\bar z(1-\bar z))^{-2\bar h}{\cal F}_{\partial\cal O}(z,\bar z),
\ee
where once again ${\cal F}_{\partial\cal O}$ is regular and its relationship to ${\cal F}_{\cal O}$ can be found in Eqs.  \eqref{F derivative} and \eqref{F derivative coefficients}. Using these expressions, \eqref{Delta G} becomes
\begin{align}
\exp\left(-\Delta S_A^{(2)}(z,\bar z)\right)=\frac{1}{4 h^2 (1 + 2 h)^2}\left[ {\cal F}_{\partial\cal O}(z,\bar z)+\tilde\Xi (z,\bar z)\right],
\end{align}
where we introduced
\begin{equation}
\tilde\Xi (z,\bar z)=((1 - z) z)^{2 (1 + h)} ((1 -\bar z)\bar  z)^{2\bar h}\,\Xi (z,\bar z)\,.
\end{equation}

\medskip
Now, we calculate $\Delta S_A^{(2)}(z,\bar z)$ after removing the cut-off. As discussed in section \ref{sec:replica} we have to possibilities, either $t\in [l,L+l]$ or $t\notin [l,L+l]$, for which the cross ratios become $(z,\bar z)\simeq (1,0)$ and $(z,\bar z)\simeq (0,0)$ respectively. 
Expanding $\tilde\Xi$ about the former yields
\begin{align}
\tilde\Xi (z,\bar z)\simeq&(1- z)\,\bigg[ \frac{4h-2 }{(1 + 2 h)^2}\, {\cal F}_{\cal O}(1,0)+\left(\frac{1 - 2 h}{1 + 2 h} \right)^2 \partial  {\cal F}_{\cal O}(1,0)\bigg] + \bar z\,\bar\partial {\cal F}_{\cal O}(1,0)+\,\dots\,.
\end{align}
while for the latter we find
\begin{align}
\tilde\Xi (z,\bar z)\simeq& z\,\bigg[ \frac{4h-2 }{(1 + 2 h)^2}\, {\cal F}_{\cal O}(0,0)+\left(\frac{1 - 2 h}{1 + 2 h} \right)^2 \partial  {\cal F}_{\cal O}(0,0)\bigg] + \bar z\,\bar\partial {\cal F}_{\cal O}(0,0)+\,\dots
\end{align}
Thus, both of these contributions vanish. Finally, using Eq. \eqref{F derivative} we find 
\be
\Delta S^{(2)}_A\simeq \left\{
\begin{array}{ll}
 -\log {\cal F}_{\cal O}(0,0)&\qquad t\notin [l, L+l] \\
 -\log {\cal F}_{\cal O}(1,0) &\qquad t\in [l, L+l] \label{2Rex1}
\end{array}
\right.
\ee
which coincides with the result obtained for primary operators. The fact that we are considering a descendant operator instead of a 
a primary has a tangible effect. Namely, the total energy injected into the system is divided in different proportions amongst the left and right movers. In fact we would expect the action of  each $L_{-n}$ ($\bar L_{-n}$) to increase the height of the right (left) moving lump. Notice, however, that this unbalance doesn't affect the jump in the second R\'enyi entropy of the system. It is only important that we have a pair of lumps propagating in the opposite directions and that one of the members of the pair is inside the entangling region of an arbitrary shape (in 1d either a finite or semi-infinite interval) while the other member remains outside. This strongly hints to the topological nature of this quantity (see also \cite{Dong:2008ft} and the discussion section).

%%%%%%%%%%%%%%%%%%%%%%%%%%
\subsection{Both derivatives}
%%%%%%%%%%%%%%%%%%%%%%%%%%

\begin{figure}[h!]	
  \centering
  \includegraphics[width=11cm]{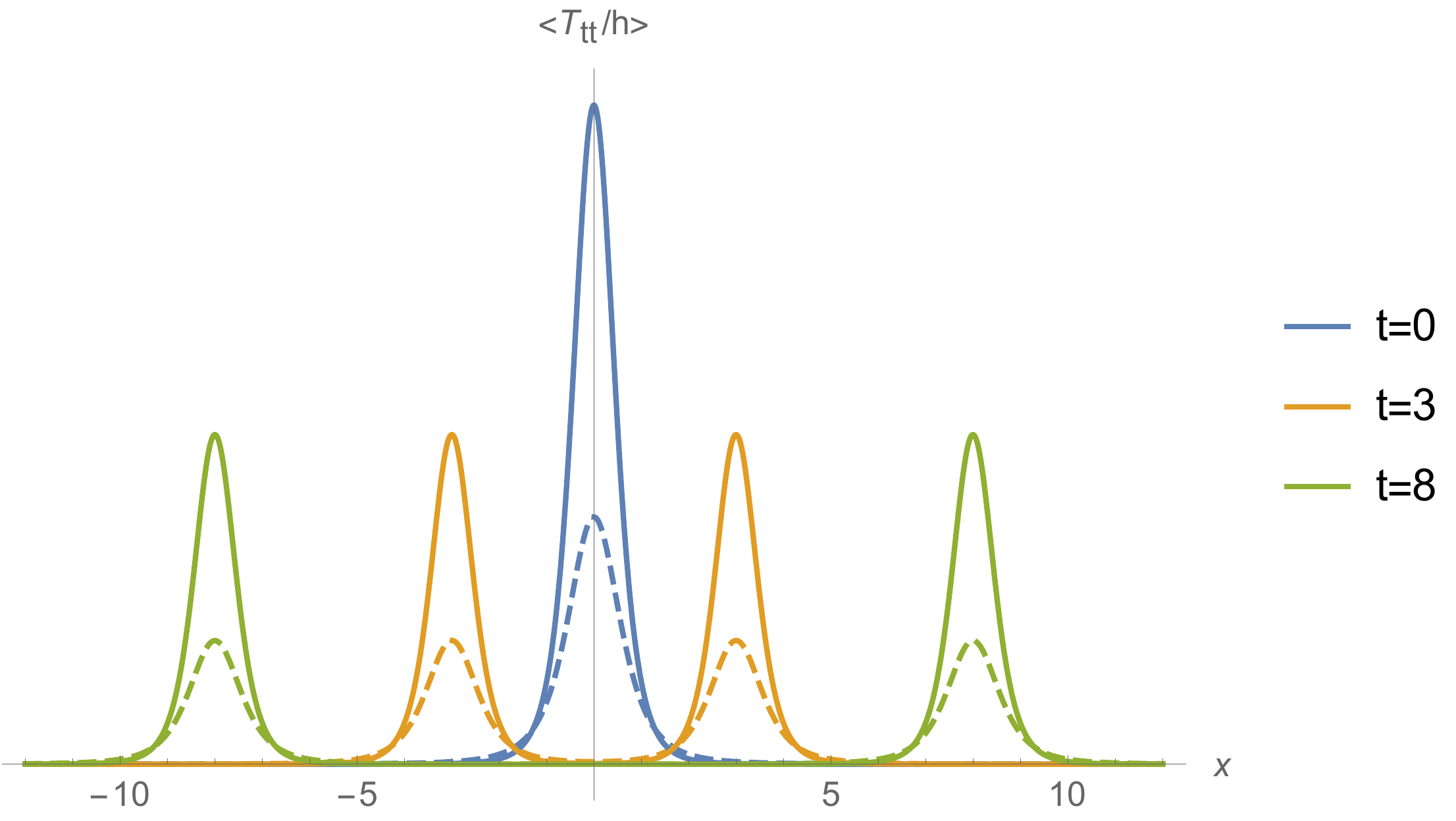}
  \caption{Evolution of the energy density for three different times in the state excited by $\bar{\partial}\partial O$. The dashed lines show the evolution in the state excited by a primary. Plot for $l=0$ and $\epsilon=1$.}\label{EnergyEx2d}
\end{figure}

In this section, we briefly consider the change in the second R\'enyi entropy due to the insertion of the second order descendant 
\be
\bar\partial\partial{\cal O}(z,\bar{z})=\bar L_{-1}L_{-1}{\cal O}(z,\bar{z})= \partial_{\bar z} \partial_z {\cal O}(z,\bar{z})\,.\label{Desa2}
\ee
The energy density profile corresponding to this descendant is given by 
\begin{align}
\langle T_{tt}(x,x)\rangle_{\bar{\partial}\partial {\cal O}}=&\langle T_{tt}(x,x)\rangle_{{\cal O}}+\frac{4\epsilon^2\left[(1-4h)(x+l-t)^2+(1+4h)\epsilon^2\right]}{(2h+1)\left[(x+l-t)^2+\epsilon^2\right]^3}\nn\\
&+\frac{4\epsilon^2\left[(1-4\bar{h})(x+l+t)^2+(1+4\bar{h})\epsilon^2\right]}{(2\bar{h}+1)\left[(x+l+t)^2+\epsilon^2\right]^3}\,,
\end{align}
and its time evolution is depicted in Fig. \ref{EnergyEx2d}. Moreover, the total energy injected into the system is 
\be 
E_{\bar\partial\partial{\cal O}}=\frac{2\pi(h+\bar{h}+2)}{\epsilon}\,.
\ee
Notice that the left/right symmetry of the energy carried by the wave packets is restored.  

\medskip
The task of calculating the $\Delta S_A^{(2)}$ corresponding to the insertion of $\bar\partial\partial{\cal O}$ is analogous to those performed in the previous sections. Nevertheless, there are some complications that we wish to point out. First, the transformation of the operator $\bar\partial\partial{\cal O}$ under conformal maps is more complicated, it contains contributions proportional to 
$\bar\partial\partial{\cal O}$, $\partial{\cal O}$ and ${\cal O}$. The explicit form of this transformation can be found in Eq. \eqref{desc trans}. Once again, we find an expression of the form
\begin{align}
\exp\left(-\Delta S_A^{(2)}(z,\bar z)\right)={\cal N}_{\bar\partial\partial{\cal O}}\left[ ~{\cal G}_{\bar\partial\partial{\cal O}}(z,\bar z)+\dots\right]\,,
\end{align}
where the dots contain the elements of lower order in derivatives. Furthermore, the normalization factor reads 
\be
{\cal N}_{\bar\partial{\cal O}\partial{\cal O}}=\left(\frac{((1 - z) z)^{2 (1 + h)}}{4 h^2 (1 + 2 h)^2}\right) \left(\frac{ ((1 -\bar z)\bar  z)^{2(1+\bar h)}}{4\bar h^2 (1 + 2\bar h)^2}\right).
\ee
As well, the function ${\cal G}_{\bar\partial\partial{\cal O}}$ becomes more convoluted; nevertheless, it can be expressed in terms ${\cal G}_{{\cal O}}$, see Eq. \eqref{G dbar d}. Despite these complications, we can show that 
\begin{align}
\exp\left(-\Delta S_A^{(2)}(z,\bar z)\right)=\left( 4 h (1 + 2 h) \bar h (1 + 2\bar h)\right)^{-2}\left[ ~{\cal F}_{\bar\partial\partial{\cal O}}(z,\bar z)+\dots\right]\,,
\end{align}
where ${\cal F}_{\bar\partial\partial{\cal O}}$ is given by \eqref{fdd}. Finally, as we remove the UV cut-off we find that
\be
\Delta S^{(2)}_A\simeq \left\{
\begin{array}{ll}
 -\log {\cal F}_{\cal O}(0,0)&\qquad t\notin [l, L+l] \\
 -\log {\cal F}_{\cal O}(1,0) &\qquad t\in [l, L+l] \label{2Rex2}
\end{array}
\right.
\ee
once more. This clearly shows that the propagation of the energy in the state excited by the descendants only provides a qualitative support for the quasi-particle picture.

%%%%%%%%%%%%%%%%%%%%%%%%%%
\subsection{Descendants at level 2}\label{DesL2}
%%%%%%%%%%%%%%%%%%%%%%%%%%

It is always possible to write the correlators of descendants in terms of the action of certain differential operators on the correlator of the primary operators (family's parent). Whenever the correlator in question involves only (powers of) $\hat{L}_{-1}$ and $\hat{\bar L}_{-1}$ the Virasoro generators can be pulled out of the correlator as simple derivatives. However, if there is any descendant constructed with an $\hat{L}_{-k}$ with $k>1$ the situation changes. For example, if there is one such descendant in the correlator then \cite{DiF}
\be
\langle{\cal O}(w_1,\bar w_1)\dots\hat  L_{-k}{\cal O}(w_i,\bar w_i)\dots {\cal O}(w_N, \bar w_N)\rangle=\mathcal{L}^{(i)}_{-k}\langle{\cal O}(w_1,\bar w_1)\dots{\cal O}(w_N, \bar w_N)\rangle\,,
\ee
where
\be
\mathcal{L}^{(i)}_{-k}=\sum_{j\neq i}\left(\frac{h(k-1)}{w^k_{ji}}-\frac{\partial_j}{w_{ji}^{k-1}}\right).
\ee
However, if the correlator contains more than one descendant the appropriate commutation of the Virasoro algebras at different points must be taken into account. This is manifest already for level $k=2$ and we discuss it below.

\medskip
In the following we compute $\Delta S_A^{(2)}$ due to the insertion of the operator
\be
 \mathcal{O}^{(-2)}(z,\bar{z})=\hat{L}_{-2}\mathcal{O}(z,\bar{z})\,.
\ee
The crucial correlation function that will contain the relevant constant is the four point function of $ \mathcal{O}^{(-2)}$, which appears in the numerator of  Eq. \eqref{eq:Delta}\footnote{In the limit of $(z,\bar{z})\to(1,0)$ all the other correlators are suppressed by powers of $(1-z)$}. We proceed in a standard way \cite{DiF}, first we write the correlator as
\be
\langle\hat L_{-2}\mathcal{O}_1\mathcal{O}^{(-2)}_2\mathcal{O}^{(-2)}_3\mathcal{O}^{(-2)}_4\rangle=\sum_{i=2}^4\oint_{{\cal C}(w_i)}\frac{dz}{2\pi i} (z_1-z)^{-1}\langle\mathcal{O}_1\dots\left( T(z)\mathcal{O}^{(-2)}_i\right)\dots\mathcal{O}^{(-2)}_4\rangle \,,\label{eq:first step}
\ee
and then make use of the OPE 
\be
T(z)\mathcal{O}^{(-2)}(w)\sim \frac{\left(\frac{c}{2}+4h\right)\mathcal{O}(w)}{(z-w)^4}+\frac{3\partial\mathcal{O}(w)}{(z-w)^3}+\frac{(h+2)\mathcal{O}^{(-2)}(w)}{(z-w)^2}+\frac{\partial\mathcal{O}^{(-2)}(w)}{z-w}. \label{OPE}
\ee
Inserting this OPE into \eqref{eq:first step} and calculating the residues, we can express the four point function in terms of correlators containing one primary and three descendants. Repeating this procedure, it is possible to reduce the number of descendants in the RHS until the full correlator is written in terms of the four-point function of the primary operators. The full procedure is rather cumbersome and the result is given by \eqref{Four point function -2}.

\medskip
Once more, after insertion $z_3=-z_1$ and $z_4=-z_2$, we define
\be
\langle\mathcal{O}^{(-2)}_1\mathcal{O}^{(-2)}_2\mathcal{O}^{(-2)}_3\mathcal{O}^{(-2)}_4\rangle=\left(4z_{1}z_{2}\right)^{-2(h+1)}\left(4\bar z_{1}\bar z_{2}\right)^{-2 \bar h} {\cal G}_{{\cal O}^{(-2)}}(z,\bar z),
\ee
with $\mathcal{O}_i=\mathcal{O}(z_i,\bar{z}_i)$, as well as 
\be
{\cal G}_{{\cal O}^{(-2)}}(z,\bar z)=\left(z(1-z)\right)^{-2(h+1)}\left(\bar z(1-\bar z)\right)^{-2 \bar h}{\cal F}_{{\cal O}^{(-2)}}(z,\bar z)\,.
\ee
The increase on the second R\'enyi entropy can be written as 
\begin{align}
\exp\left(-\Delta S_A^{(2)}(z,\bar z)\right)=&{\cal N}_{{O}^{(-2)}}\left[ ~{\cal G}_{{O}^{(-2)}}(z,\bar z)+\dots\right]\,,
\end{align}
where the pre-factor reads
\be
{\cal N}_{{O}^{(-2)}}=\frac{ \left(4z_{1}z_{2}\right)^{-2(h+2)}\left(4\bar z_{1}\bar z_{2}\right)^{-2 \bar h}\prod_{i=1}^4 \alpha_i\bar\alpha_i}{\left(\langle\mathcal{O}^{(-2)}(w_1,\bar{w}_1)\mathcal{O}^{(-2)}(w_2,\bar{w}_2)\rangle_{\Sigma_1} \right)^2}\,.
\ee
Using the OPE \eqref{OPE}, we can show that the two-point function is given by 
\be
\langle\mathcal{O}^{(-2)}(w_1,\bar{w}_1)\mathcal{O}^{(-2)}(w_2,\bar{w}_2)\rangle_{\Sigma_1} =\frac{1}{2} \left(c + 2 h (9 h + 22)\right)w_{12}^{-2(h+2)}\bar w_{12}^{-2h}.
\ee
Finally, we extract ${\cal F}_{{\cal O}^{(-2)}}$ from Eq. \eqref{Four point function -2}, and as we remove the UV cut-off $\epsilon$ find
\be
{\cal F}_{{O}^{(-2)}}(z,\bar z)\simeq \frac{1}{4} \left(c + 2 h (9 h + 22)\right)^2 \left\{
\begin{array}{ll}
 {\cal F}_{\cal O}(0,0)&\qquad t\notin [l, L+l] \\
 {\cal F}_{\cal O}(1,0) &\qquad t\in [l, L+l] \label{2Rex3}
\end{array}
\right.
\ee
Therefore, we find that for the insertion of the operator ${{O}^{(-2)}}$ we have
\be
\Delta S^{(2)}_A\simeq \left\{
\begin{array}{ll}
 -\log {\cal F}_{\cal O}(0,0)&\qquad t\notin [l, L+l] \\
 -\log {\cal F}_{\cal O}(1,0) &\qquad t\in [l, L+l] 
\end{array}
\right. 
\ee
as expected.

\medskip

Clearly the procedure becomes more involved once the level of the descendant increases but the increase in the Renyi entropy remains fixed by the two limits of the original ${\cal F}_{\cal O}(z,\bar{z})$ from the correlator of the primary operators.

%%%%%%%%%%%%%%%%%%%%%%%%%%
\section{General derivation}
%%%%%%%%%%%%%%%%%%%%%%%%%%

Based on the explicit examples in the previous sections, we can clearly deduce that the relevant contribution to the R\'enyi entanglement entropies is the same for the whole conformal family. We show it for the second R\'enyi entropy but the generalization to higher $n$ should be straightforward (though very involved computationally).

\medskip
 The technical reason for the constant is the following\footnote{Similar logic has been applied in \cite{Brustein:1988vb}}. On one hand, under conformal transformation, every descendant of level $(n,\bar{n})$ transforms as
\be
{\cal O}^{(n,\bar{n})}(z,\bar{z})=\left(f'\right)^{h+n}\left(\bar{f}'\right)^{\bar{h}+\bar{n}}\left[{\cal O}^{(n,\bar{n})}(f,\bar{f})+...\right]\label{trd}
\ee
where the ellipsis stand operators of the lower dimension that in the correlation functions give rise to lower order singularity. This way, once we employ the map \eqref{map}, the Jacobian from \eqref{trd} with the square of the two-point function on $\Sigma_1$ \eqref{eq:Delta} combine into
\be
P=\mathcal{N}^{-2}_2[4z_1z_2]^{h+n}[4\bar{z}_1\bar{z}_2]^{\bar{h}+\bar{n}}[z(1-z)]^{h+n}[\bar{z}(1-\bar{z})]^{\bar{h}+\bar{n}},
 \ee
with $\mathcal{N}_2$ being the constant in the two-point function of the descendants.

\medskip
 When we extract the constant contribution to the entanglement R\'enyi entropy we take the limit of $(z,\bar{z})\to (1,0)$. The only terms that survive this regime must cancel the above pre-factor and are confined to the correlation function of $(n,\bar{n})$ descendants only.\\
 On the other hand, the correlators of descendants on the plane can be obtained from the correlators of the primary operators that we define as \footnote{This is a form that we assume with $F(z,\bar{z})$ being smooth at $z\to1$ as well as $z\to0$. This is holds for all the cases that we are aware of.}
\be
\langle {\cal O}_{\alpha}(z_1,\bar{z}_1){\cal O}^\dagger_{\alpha}(z_2,\bar{z}_2){\cal O}_{\alpha}(z_3,\bar{z}_3){\cal O}^\dagger_{\alpha}(z_4,\bar{z}_4)\rangle\equiv |z_{13}z_{24}|^{-4h}|z(1-z)|^{-4h}\mathcal{F}_{\alpha}(z,\bar{z}),
\ee
by acting with a (complicated but in principle straightforward to derive) differential operator. In the limit of $(z,\bar{z})\to (1,0)$ the only contribution that has enough singularity to cancel the above-mentioned pre-factor comes from the terms in which the differential operator leaves $\mathcal{F}_{\alpha}(z,\bar{z})$ intact and contains precisely the square of the norm $\mathcal{N}_{{\cal O}^{(n,\bar{n})}}$ of the two-point function\footnote{We have verified it up to level 2 and for arbitrary powers of derivatives}.\\ 
This way the non-trivial constant obtained from the correlators in the limit of $(z,\bar{z})\to(1,0)$ is the same for all the members of a given conformal family. By the same token, one can show that  as $(z,\bar{z})\to(0,0)$ the R\'enyi entanglement entropy is unchanged.

%%%%%%%%%%%%%%%%%%%%%%%%%%
\section{Conclusions and discussion}
%%%%%%%%%%%%%%%%%%%%%%%%%%

In this work we have demonstrated that, in 1+1 dimensional CFTs with Virasoro symmetry, local excitations by operators in the same conformal family increase the second R\'enyi entanglement entropy by the same constant. For rational conformal field theories this entanglement constant is the logarithm of the quantum dimension (obtained in \cite{He:2014mwa}) of the primary operator that represents the family.
We have checked this on explicit examples and outlined a general derivation. A generalization to higher R\'enyi entropies should be also possible and we expect it to work in exactly the same way as for the second R\'enyi entropy\footnote{Even though general 2n-point correlators of descendants will have a very complicated structure, the late time behaviour of the invariant cross-ratios is universal. Therefore, at least in RCFTs, one could possibly use the factorisation and the fusion transformation for any members of a conformal family to formally demonstrate our claims for general $n$. The same should be true for $n<1$ and it would be interesting to check it for the logarithmic negativity (see e.g. \cite{Calabrese:2012ew,Wen:2015qwa})}.

\medskip
Our results strongly support the existence of universal features of entanglement for locally excited states. So far only the entanglement entropy of a block in the ground state of a local hamiltonian has been used to fix the central charge of the CFT that govern the critical points. It would be interesting to perform numerical study of the evolution of the entanglement entropy in a chain (e.g. Ising chain) excited by a local operator. Similarly to \cite{Stojevic:2014zta}, by tuning the parameters close to criticality\footnote{Generically, local operators on the chain, at the critical point, will correspond to primaries, descendants or the linear combinations of thereof.} it should be possible to verify the CFT prediction for the constant contribution to the entropy \cite{WIP}.

\medskip
Let us also discuss two different perspectives that we hope can lead to a simpler proof of our result and shed more light on the physical meaning of the constant contribution to the R\'enyi entanglement entropies.

\medskip
We have stressed that the constant contribution to the R\'enyi entropies is equivalent to the entropy between Left and Right (chiral and anti-chiral) movers in a quantum state \cite{Nozaki:2014hna} 
\be
\ket{\psi_{LR}}={\cal O}(z,\bar{z})\ket{0}_L\ket{0}_R\label{EST}
\ee
Formally, this means that we can write the state in the Schmidt form
\be
\ket{\psi_{LR}}=\sum_i\sqrt{p_i}\ket{i_L}\ket{i_R}
\ee
with Schmidt  bases $\ket{i_L}$ for $L$ and $\ket{i_R}$ $R$ and Schmidt coefficients satisfying
\be
\sum_i p_i=1
\ee
where number of non-zero $p_i$s is called the Schmidt number of state $\ket{\psi_{LR}}$ \cite{NCH}.
This way the constant contribution to the $n$-th R\'enyi entropy can be computed as
\be
\Delta S^{(n)}=\frac{1}{1-n}\log\left(\sum_{i} p^n_i\right)\label{Dsn}
\ee
and comparing \eqref{Dsn} and \eqref{DS2r} we get for rational CFTs: $F_{00}[\mathcal{O}]=\sum_{i} p^2_i$.

\medskip
The fact that this number is the same for all the members of the family of a give descendant indicates that only the elements of the Schmidt basis $\ket{i_L}$ $(\ket{i_R})$ can change from one descendant to another however the Schmidt coefficients remain the same.  

\medskip
In principle one should be able to express the excited state \eqref{EST} in the left-right orthonormal basis of descendants and then transform to the Schmidt basis using the singular value decomposition. This can be done for some excited states that have a natural left-right decomposition (see \cite{Miyaji:2014mca,PandoZayas:2014wsa,Das:2015oha,Brehm:2015lja}) but it would be interesting to preform it for \eqref{EST} even in known RCFTs.

\medskip
Let us also point that the logarithm of quantum dimension appears naturally in the computation of topological entanglement entropy in 2+1 dimensions. In fact it was shown in \cite{Dong:2008ft} that for a given region $A$ the increase of the topological entanglement entropy due to excitation $a$ is equal 
\be
\Delta S_{top}=\log (d_a)
\ee
where $d_a$ is the quantum dimension of the excitation.

\medskip
Even though rational CFT are naturally linked to topological field theories in 2+1 dimensions (e.g. FQHe),  the precise link to the 2d CFT technology employed above is not obvious. However, based on the connection between the topological entropy and the boundary entropy \cite{Fendley:2006gr}, or left-right entropy and TQFT in 2+1 \cite{Das:2015oha} \footnote{See also \cite{McGough:2013gka} for an interesting application to black holes in $AdS_3$}, it is not unlikely that the constant contribution to R\'enyi entanglement entropies in 1+1 d CFT is equivalent to the change in topological entanglement entropy in 2+1d.\\
If this happens to be the case, our results for descendants\footnote{That give rise to wave functions for excited states (see e.g.\cite{Herwerth:2015pga})} imply that the contribution from the whole topological sector (conformal family) is the same. It would be very interesting to sharpen this connection and we leave it for the future work.  

%%%%%%%%%%%%%%%%%%%%%%%%%%%%%%%
\subsection*{Acknowledgements}
%%%%%%%%%%%%%%%%%%%%%%%%%%%%%%%

We would like to thank Diptarka Das, Leszek Hadasz, Masahiro Nozaki, Konstantin Zarembo for discussions and Tadashi Takayanagi, Masahiro Nozaki, Tokiro Numasawa and Kento Watanabe for comments on the draft. We would like to thank the Galileo Galilei Institute for Theoretical Physics for the hospitality and the INFN for partial support where this work was initiated. PC is supported by  the Swedish Research Council (VR) grant 2013-4329.  The research of AVO is supported by the University Research Council of the University of the Witwatersrand. AVO wishes to thank the visitors program of the HECAP section at ICTP for their support.

%%%%%%%%%%%%%%%%%%%%%%%%%%%%%%%%%%%%%%%
%%%%%%%%%%%%%%%%%%%%%%%%%%%%%%%%%%%%%%%
%%%%%%%%%%%%%%%%%%%%%%%%%%%%%%%%%%%%%%%
\begin{appendix}
%%%%%%%%%%%%%%%%%%%%%%%%%%%%%%%%%%%%%%%
%%%%%%%%%%%%%%%%%%%%%%%%%%%%%%%%%%%%%%%
%%%%%%%%%%%%%%%%%%%%%%%%%%%%%%%%%%%%%%%

%%%%%%%%%%%%%%%%%%%%%%%%%%
%%%%%%%%%%%%%%%%%%%%%%%%%%
\section{Correlators of the EPR operators} \label{App:CO}
%%%%%%%%%%%%%%%%%%%%%%%%%%
%%%%%%%%%%%%%%%%%%%%%%%%%%

We have used the EPR operators
\bea
O_{\alpha}(z_i,\bar{z}_i)=\sqrt{a}\,e^{i\sqrt{2}\alpha\,\phi(z_i,\bar{z}_i)}\pm\sqrt{1-a}\,e^{-i\sqrt{2}\alpha\,\phi(z_i,\bar{z}_i)}\equiv\sqrt{a}\,O^{(+)}_i\pm\sqrt{1-a}\,O^{(-)}_i\nn\\
O^\dagger_{\alpha}(z_i,\bar{z}_i)=\sqrt{a}\,e^{-i\sqrt{2}\alpha\,\phi(z_i,\bar{z}_i)}\pm\sqrt{1-a}\,e^{i\sqrt{2}\alpha\,\phi(z_i,\bar{z}_i)}\equiv\sqrt{a}\,O^{(-)}_i\pm\sqrt{1-a}\,O^{(+)}_i  \label{Opa}
\eea
with conformal dimension $h=\bar{h}=\alpha^2$ and parameter $a\in[0,1]$. \\
Here we compute the four-point correlator on the complex plane
\be
G^{4}_{O_\alpha}=\langle O_{\alpha}(z_1,\bar{z}_1)O^\dagger_{\alpha}(z_2,\bar{z}_2)O_{\alpha}(z_3,\bar{z}_3)O^\dagger_{\alpha}(z_4,\bar{z}_4)\rangle
\ee
Since the non zero correlators of the vertex operators are the ones that have the zero sum of the operators dimensions
\be
\langle\prod_ie^{i\sqrt{2}\alpha_i\,\phi(z_i,\bar{z}_i)}\rangle=\prod_{i<j}|z_i-z_j|^{4\alpha_i\alpha_j},\qquad \sum_i\alpha_i=0 
\ee
we denote
\be
G^{(a_1,a_2,a_3,a_4)}\equiv\langle O^{(a_1)}_1O^{(a_2)}_2O^{(a_3)}_3O^{(a_4)}_4\rangle,\qquad a_i\in \{+,-\}
\ee 
And the non-zero contributions to the correlator comes from
\be
G^{4}_{O_\alpha}=(a^2+(1-a)^2)G^{(+,-,+,-)}+2a(1-a)\left(G^{(+,+,-,-)}+G^{(+,-,-,+)}\right)
\ee
where we used
\be
G^{(+,-,+,-)}=G^{(-,+,-,+)},\qquad G^{(+,+,-,-)}=G^{(-,-,+,+)},\qquad G^{(+,-,-,+)}=G^{(-,+,+,-)}
\ee
The remaining correlators are
\bea
G^{(+,-,+,-)}&=&\frac{1}{|z_{13}z_{24}|^{4h}|z(1-z)|^{4h}}\nn\\
G^{(+,+,-,-)}&=&\frac{|z|^{8h}}{|z_{13}z_{24}|^{4h}|z(1-z)|^{4h}}\nn\\
G^{(+,-,-,+)}&=&\frac{|1-z|^{8h}}{|z_{13}z_{24}|^{4h}|z(1-z)|^{4h}}
\eea
The four-point correlator of such operators on the complex plane is given by
\be
G^{4}_{O_\alpha}=|z_{13}z_{24}|^{-4h}{\cal G}(z,\bar{z})
\ee
with the canonical 4-pt function 
\be
{\cal G}(z,\bar{z})\equiv \langle O_\alpha|O_{\alpha}(z,\bar{z})O_{\alpha}(1,1)|O_{\alpha}\rangle\equiv|z(1-z)|^{-4h}\mathcal{F}(z,\bar{z})
\ee
where 
\be
\mathcal{F}(z,\bar{z})=a^2+(1-a)^2+2a(1-a)\left(|z|^{8h}+|1-z|^{8h}\right).
\ee

%%%%%%%%%%%%%%%%%%%%%%%%%%
%%%%%%%%%%%%%%%%%%%%%%%%%%
\section{Transformation of descendants} \label{App:Des}
%%%%%%%%%%%%%%%%%%%%%%%%%%
%%%%%%%%%%%%%%%%%%%%%%%%%%

The transformation of descendant fields $\phi^{(-k)}$ (of a primary $\phi$ with dimension $h$) under finite conformal map $f(z)$ has been worked out in \cite{Gaberdiel:1994fs} (see also \cite{Palmai:2014jqa}). For the first few descendants it can be obtained from
\be
\phi^{(-k)}(z)=e^{R_0 L_0}\left(1+R_1L_1+\frac{1}{2}R^2_1L^2_1+R_2L_2+...\right)\phi^{(-k)}(f)\nn
\ee
where
\bea
R_0=\log f',\qquad R_1=\frac{1}{2}R'_0=\frac{f''}{2f'},\qquad R_2=\frac{Sf}{6} \nn
\eea
where $Sf$ is the Schwarzian derivative of f. Using the Virasoro algebra and the commutation relations of $L$'s
with the operator modes $\phi_n$ we can derive
\bea
\phi(z)&=&\left(f'\right)^{h}\phi(f)\nn\\
\phi^{(-1)}(z)&=&\left(f'\right)^{h+1}\left[\phi^{(-1)}(f)+h\frac{f''}{f'^2}\phi(f)\right]\nn\\
\phi^{(-1,-1)}(z)&=&\left(f'\right)^{h+2}\left[\phi^{(-1,-1)}(f)+(2h+1)\frac{f''}{f'^2}\phi^{(-1)}(f)+h\left((2h+1)\frac{f''^2}{2f'^4}+\frac{Sf}{f'^2}\right)\phi(f)\right]\nn\\
\phi^{(-2)}(z)&=&\left(f'\right)^{h+2}\left[\phi^{(-2)}(f)+\frac{3f''}{2f'^2}\phi^{(-1)}(f)+\left(\frac{3hf''^2}{4f'^4}+\left(4h+\frac{c}{2}\right)\frac{Sf}{6f'^2}\right)\phi(f)\right]\nn\\ \label{TrDes}
\eea 
The first two descendants correspond to derivatives and the transformation rule can be also reproduced from
\be
\phi^{(-1)}(z)=\partial_z\left[\left(f'\right)^{h}\phi(f)\right],\qquad \phi^{(-1,-1)}(z)=\partial^2_z\left[\left(f'\right)^{h}\phi(f)\right]\nn
\ee 
It is also easy to check that for $h=0$ (the identity) the transformation of $\phi^{(-2)}(z)$ reproduces that standard transformation for $T(z)$. \\
We have also used the transformations
\bea
\partial\bar{\partial}\phi(z,\bar{z})=(f')^{h+1}(\bar{f}')^{\bar{h}+1}\left[\partial\bar{\partial}\phi(f,\bar{f})+h\frac{f''}{f'^2}\bar{\partial}\phi(f,\bar{f})+\bar{h}\frac{\bar{f}''}{\bar{f}'^2}\partial\phi(f,\bar{f})+h\bar{h}\frac{f''}{f'^2}\frac{\bar{f}''}{\bar{f}'^2}\phi(f,\bar{f})\right]\nn\\
\partial\bar{\partial}\phi(f,\bar{f})=(f')^{-(h+1)}(\bar{f}')^{-(\bar{h}+1)}\left[\partial\bar{\partial}\phi(z,\bar{z})-h\frac{f''}{f'}\bar{\partial}\phi(z,\bar{z})-\bar{h}\frac{\bar{f}''}{\bar{f}'}\partial\phi(z,\bar{z})+h\bar{h}\frac{f''}{f'}\frac{\bar{f}''}{\bar{f}'}\phi(z,\bar{z})\right]\nn\\ \label{desc trans}
\eea

%%%%%%%%%%%%%%%%%%%%%%%%%%
%%%%%%%%%%%%%%%%%%%%%%%%%%
\section{Correlators at level 1} \label{App:CB}
%%%%%%%%%%%%%%%%%%%%%%%%%%
%%%%%%%%%%%%%%%%%%%%%%%%%%

In the following, we display some results concerning the four-point function of the operator $\partial{\ca O}$
on the plane.  If we define 
\be
  \langle {\cal O}(z_1,\bar z_1){\cal O}(z_2,\bar z_2){\cal O}(z_3,\bar z_3){\cal O}(z_4,\bar z_4)\rangle=\left(z_{13}z_{24}\right)^{-2h}\left(\bar z_{13}\bar z_{24}\right)^{-2 \bar h} {\cal G}_{\cal O}(z,\bar z),
\ee
as well as\footnote{This form assumes $z_3=-z_1$ and $z_4=-z_2$ }
\be
  \langle\partial {\cal O}(z_1,\bar z_1)\partial{\cal O}(z_2,\bar z_2)\partial{\cal O}(z_3,\bar z_3)\partial{\cal O}(z_4,\bar z_4)\rangle=\left(z_{13}z_{24}\right)^{-2(h+1)}\left(\bar z_{13}\bar z_{24}\right)^{-2 \bar h} {\cal G}_{\partial\cal O}(z,\bar z)\,,
\ee
we can show that ${\cal G}_{\cal O}(z,\bar z)$ and ${\cal G}_{\partial\cal O}(z,\bar z)$ are related by
\be
{\cal G}_{\partial\cal O}(z,\bar z)= \sum_{l=0}^{4}g_l (z)\;\partial^l{\cal G}_{\cal O}(z,\bar z)\, \label{G derivative}
\ee
where
\begin{align}
g_0 (z)&=4 h^2 (1 + 2 h)^2\,,\nonumber\\
g_1 (z)&=2 (1 + 2 h) ( 2 z-1)\,,\nonumber\\
g_2 (z)&=2 + 2 z (4 h^2-7 - 2 h ) (1 - z) \,,\nonumber\\
g_3 (z)&=-4\, z (1-z) (2z-1)\nonumber\,,\\
g_4 (z)&=(1 - z)^2 z^2 \;. \label{G derivative coefficients}
\end{align}
Moreover, if we define the functions
\be
{\cal F}_{\cal O}(z,\bar z)=(z(1-z))^{2h}(\bar z(1-\bar z))^{2\bar h}{\cal G}_{\cal O}(z,\bar z)\;,
\ee
and
\be
{\cal F}_{\partial\cal O}(z,\bar z)=(z(1-z))^{2(h+1)}(\bar z(1-\bar z))^{2\bar h}{\cal G}_{\partial\cal O}(z,\bar z)\;,
\ee
we can show that
\be
{\cal F}_{\partial\cal O}(z,\bar z)= \sum_{l=0}^{4}f_l (z)\; \partial^l {\cal F}_{\cal O}(z,\bar z)\;, \label{F derivative}
\ee
where
\begin{align}
f_0(z)=&4 h^2 [(1 + (-1 + z) z)^2 + 4 h^2 (1 + 3 (-1 + z) z)^2\nonumber\\ - 
   &4 h (-1 + (-1 + z) z (-2 + 5 (-1 + z) z))]\nonumber\\
f_1(z)=&-2 (-1 + z) z (-1 + 2 z) (z + 10 h (-1 + z) z-  z^2 \nonumber\\
& - 40 h^2 (-1 + z) z + 16 h^3 (1 + 3 (-1 + z) z))\nonumber\\
f_2(z)=&2 (-1 + z)^2 z^2 (1 + 7 (-1 + z) z + h (-6 - 34 (-1 + z) z) \nonumber\\
&+4 h^2 (3 + 11 (-1 + z) z))\nonumber\\
f_3(z)=&-4 (-1 + 2 h) (-1 + z)^3 z^3 (-1 + 2 z)\nonumber\\
f_4(z)=&(-1 + z)^4 z^4\; . \label{F derivative coefficients}
\end{align}

Note: At the points $(z,\bar z)\rightarrow (0,0),\,(1,0)$, the only contribution to ${\cal F}_{\partial\cal O}(z,\bar z)$ that survives comes from the $f_0(z)$ term in \eqref{F derivative}, which in terms of the coefficients of \eqref{G derivative} reads
\be
{\cal F}_{\partial\cal O}(z,\bar z)\rightarrow 2 h (1 + 2 h) \left[g_2(z) + 2 (1 + h) ((3 + 2 h) \gamma_4(z)-\gamma_3 (z))\right]  {\cal F}_{\cal O}(z,\bar z),
\ee
where $g_3 (z)=z(1-z) \gamma_3(z)$ and $g_4(z)=z^2(1-z)^2 \gamma_4(z)$.

On the other hand, for the descendant $\bar\partial\partial{\cal O}$ we have
\be
{\cal G}_{\bar\partial\partial\cal O}(z,\bar z)= \sum_{k,l=0}^{4}g_{ k ,  l} (z,\bar z)\;\partial^k\bar{\partial}^l{\cal G}_{\cal O}(z,\bar z)\, \label{G dbar d}
\ee
where 
\be
g_{k,l}(z,\bar z)= g_k(z)\bar g_l (\bar z),
\ee
where the functions $g_i(z)$ are defined in \eqref{G derivative coefficients}.

\be
{\cal F}_{\bar\partial\partial\cal O}(z,\bar z)=(z(1-z))^{2(h+1)}(\bar z(1-\bar z))^{2(\bar h+1)}{\cal G}_{\bar\partial\partial\cal O}(z,\bar z)\;,\label{fdd}
\ee

%%%%%%%%%%%%%%%%%%%%%%%%%%
%%%%%%%%%%%%%%%%%%%%%%%%%%
\section{Correlators at level 2}\label{App:L2}
%%%%%%%%%%%%%%%%%%%%%%%%%%
%%%%%%%%%%%%%%%%%%%%%%%%%%

In this appendix we present some details of the computation of the correlation functions of descendants $\mathcal{O}^{(-2)}(w,\bar{w})$. The key ingredient (for any operator) is the OPE with the energy momentum tensor that in this case reads
\be
T(z)\mathcal{O}^{(-2)}(w)\sim \frac{\left(\frac{c}{2}+4h\right)\mathcal{O}(w)}{(z-w)^4}+\frac{3\partial\mathcal{O}(w)}{(z-w)^3}+\frac{(h+2)\mathcal{O}^{(-2)}(w)}{(z-w)^2}+\frac{\partial\mathcal{O}^{(-2)}(w)}{z-w}
\ee
Denoting $\mathcal{O}^{(-2)}(w_i)\equiv\mathcal{O}^{(-2)}_i$ and $\mathcal{O}(w_i)\equiv\mathcal{O}_i$ we compute on the complex plane
\bea
\langle\mathcal{O}^{(-2)}_1\mathcal{O}^{(-2)}_2\rangle&=&\left[\frac{\frac{c}{2}+4h}{w^4_{12}}+\frac{3\partial_2}{w^3_{12}}+\left(\frac{(h+2)}{w^2_{12}}+\frac{\partial_2}{w_{12}}\right)\left(\frac{h}{w^2_{12}}-\frac{\partial_1}{w_{12}}\right)\right]\langle\mathcal{O}_1\mathcal{O}_2\rangle\nn\\
&=&\frac{1}{2}\left(c+2h(9h+22)\right)w^{-2(h+2)}_{12}
\eea
where $w_{ij}=w_i-w_j$. The products of operators should be understood as acting from left to right on the correlators. When taking the contribution from the residues (see e.g. 6.6 in \cite{DiF}) there are always few ways to start evaluating the correlator and we use the convention from left to right as well.

\medskip
Analogously the three-point function can be written
\begin{align}
\langle\mathcal{O}^{(-2)}_1\mathcal{O}^{(-2)}_2\mathcal{O}^{(-2)}_3\rangle=\left[\mathcal{D}_{1,2}\mathcal{L}^{(3)}_{-2}+\mathcal{D}_{1,3}\mathcal{L}^{(2)}_{-2}+\mathcal{H}^{(1)}_{(1)}\mathcal{D}_{2,3}+\mathcal{H}^{(1)}_{(1)}\mathcal{H}^{(2)}_{(1)}
\mathcal{L}^{(3)}_{-2}\right]\langle\mathcal{O}_1\mathcal{O}_2\mathcal{O}_3\rangle\, , 
\end{align}
where
\be
\mathcal{D}_{i,j}=\frac{\frac{c}{2}+4h}{w^4_{ij}}+\frac{3\partial_j}{w^3_{ij}},\qquad\mathcal{L}^{(i)}_{-2}=\sum_{j\neq i}\left(\frac{h}{w^2_{ji}}-\frac{\partial_j}{w_{ji}}\right)
\ee
and
\be
\mathcal{H}^{(i)}_{(k)}=\sum_{j\neq i}\left(\frac{h_j}{w^2_{ij}}-\frac{\partial_j}{w_{ji}}\right),\qquad h_k=h,\quad  h_k\neq h_m=h+2.
\ee
Finally, the four point function can be written as
\begin{align}
\langle\mathcal{O}^{(-2)}_1\mathcal{O}^{(-2)}_2\mathcal{O}^{(-2)}_3\mathcal{O}^{(-2)}_4\rangle
=&\Big[\mathcal{D}_{1,2}\left(\mathcal{D}_{3,4}+\mathcal{I}^{(3)}_{(1,2)}\mathcal{L}^{(4)}_{-2}\right)+\mathcal{D}_{1,3}\left(\mathcal{D}_{2,4}+
\mathcal{I}^{(2)}_{(1,3)}\mathcal{L}^{(4)}_{-2}\right)\nn\\
+&\mathcal{D}_{1,4}\left(\mathcal{D}_{2,3}+\mathcal{I}^{(2)}_{(1,4)}\mathcal{L}^{(3)}_{-2}\right)+\mathcal{H}^{(1)}_{(1)}
\left(\mathcal{D}_{4,2}\mathcal{L}^{(3)}_{-2}+\mathcal{D}_{4,3}\mathcal{L}^{(2)}_{-2}\right)\nn\\
+& \mathcal{H}^{(1)}_{(1)}\mathcal{I}^{(4)}_{(1,4)}\left(\mathcal{D}_{2,3}+
\mathcal{I}^{(2)}_{(1,4)}\mathcal{L}^{(3)}_{-2}\right)\Big]\, \langle\mathcal{O}_1\mathcal{O}_2\mathcal{O}_3\mathcal{O}_4\rangle\,,\label{Four point function -2}
\end{align}
where
\be
\mathcal{I}^{(i)}_{(k,l)}=\sum_{j\neq i}\left(\frac{h_j}{w^2_{ij}}-\frac{\partial_j}{w_{ji}}\right),\qquad h_k=h_l=h\quad \text{and}\quad h_p=h+2 \quad \text{for}\quad p\notin\{k,l\}.
\ee
One can verify that setting $h=0$ in the above operator reproduces the four-point function \eqref{4ptT}.

%%%%%%%%%%%%%%%%%%%%%%%%%%
%%%%%%%%%%%%%%%%%%%%%%%%%%
\section{Comments on large c and holography}\label{App:Lc}
%%%%%%%%%%%%%%%%%%%%%%%%%%
%%%%%%%%%%%%%%%%%%%%%%%%%%

The late time value of the R\'enyi entanglement entropies is non-perturbative. To illustrate this fact we consider again the state excited by the descendant of identity $T(z)$. In the computation of the late time value (semi-inifinite interval) of the second R\'enyi entanglement entropy we use the \eqref{Tres}. If we first take the limit of large central charge, the contribution to the second R\'enyi entropy becomes
\be
\Delta S^{(2)}\simeq 8\log\frac{2t}{\epsilon}-2\log\frac{c}{2}
\ee
This is the logarithmic growth with time obtained in \cite{Caputa:2014vaa} (the negative part comes form the normalisation) and it should be possible to perform a similar computation for the entanglement entropy and match it with gravity using \cite{RTorig} as in \cite{Nozaki:2014uaa}.\\
 On the other hand, as we saw in the text, if we know the answer for the full correlator (all quantum gravity effects taken into account) we can first take $\epsilon\to 0$ and obtain the entanglement constant.

\end{appendix}

%%%%%%%%%%%%%%%%%%%%%%%%%%%%%%%%%%%%%%%


\begin{thebibliography}{}
%%%%%%%%%%%%%%%%%%%%%%%%%%%%%%%%%%%%%%%

\bibitem{CC}
 P.~Calabrese and J.~L.~Cardy,
  ``Entanglement entropy and quantum field theory,''
  J.\ Stat.\ Mech.\  {\bf 0406}, P06002 (2004)
  [hep-th/0405152].
  
\bibitem{cag}
P.~Calabrese and J.~L.~Cardy,
  ``Evolution of Entanglement Entropy in One-Dimensional Systems,''
  J.\ Stat.\ Mech.\  {\bf 04} (2005) P04010, cond-mat/0503393.
  
  
  \bibitem{Nozaki:2014hna}
  M.~Nozaki, T.~Numasawa and T.~Takayanagi,
  ``Quantum Entanglement of Local Operators in Conformal Field Theories,''  Phys.\ Rev.\ Lett.\  {\bf 112} (2014) 111602  [arXiv:1401.0539 [hep-th]].  %%CITATION = ARXIV:1401.0539;%%

  \bibitem{Alcaraz:2011tn}
  F.~C.~Alcaraz, M.~I.~Berganza and G.~Sierra,
  ``Entanglement of low-energy excitations in Conformal Field Theory,''
  Phys.\ Rev.\ Lett.\  {\bf 106} (2011) 201601
  [arXiv:1101.2881 [cond-mat.stat-mech]].
  %%CITATION = ARXIV:1101.2881;%%
  %\cite{Palmai:2014jqa}
\bibitem{Palmai:2014jqa} 
  T.~Palmai,
  ``Excited state entanglement in one dimensional quantum critical systems: Extensivity and the role of microscopic details,''
  Phys.\ Rev.\ B {\bf 90}, no. 16, 161404 (2014)
  [arXiv:1406.3182 [hep-th]].
  %%CITATION = ARXIV:1406.3182;%%


  \bibitem{Caputa:2014vaa} 
  P.~Caputa, M.~Nozaki and T.~Takayanagi,
  ``Entanglement of local operators in large-N conformal field theories,''
  PTEP {\bf 2014}, no. 9, 093B06 (2014)
  [arXiv:1405.5946 [hep-th]].

\bibitem{Nozaki:2014uaa}
  M.~Nozaki,
  ``Notes on Quantum Entanglement of Local Operators,''  JHEP {\bf 1410} (2014) 147  [arXiv:1405.5875 [hep-th]].  %%CITATION = ARXIV:1405.5875;%%
  
  %\cite{Shiba:2014uia}
\bibitem{Shiba:2014uia} 
  N.~Shiba,
  ``Entanglement Entropy of Disjoint Regions in Excited States : An Operator Method,''
  JHEP {\bf 1412}, 152 (2014)
  [arXiv:1408.0637 [hep-th]].
  %%CITATION = ARXIV:1408.0637;%%
  
\bibitem{He:2014mwa}
  S.~He, T.~Numasawa, T.~Takayanagi and K.~Watanabe,
  ``Quantum dimension as entanglement entropy in two dimensional conformal field theories,''
  Phys.\ Rev.\ D {\bf 90} (2014) 4,  041701
  [arXiv:1403.0702 [hep-th]].
  %%CITATION = ARXIV:1403.0702;%%
  
  \bibitem{Guo:2015uwa} 
  W.~Z.~Guo and S.~He,
  ``R�nyi entropy of locally excited states with thermal and boundary effect in 2D CFTs,''
  JHEP {\bf 1504}, 099 (2015)
  [arXiv:1501.00757 [hep-th]].
  %%CITATION = ARXIV:1501.00757;%%
  
 \bibitem{Caputa:2014eta}
  P.~Caputa, J.~Sim\'on, A.~\v{S}tikonas and T.~Takayanagi,
  ``Quantum Entanglement of Localized Excited States at Finite Temperature,''
  JHEP {\bf 1501} (2015) 102
  [arXiv:1410.2287 [hep-th]].
  
  %\cite{Asplund:2014coa}
\bibitem{Asplund:2014coa}
  C.~T.~Asplund, A.~Bernamonti, F.~Galli and T.~Hartman,
  ``Holographic Entanglement Entropy from 2d CFT: Heavy States and Local Quenches,''
  JHEP {\bf 1502} (2015) 171
  [arXiv:1410.1392 [hep-th]].
  %%CITATION = ARXIV:1410.1392;%%
  %\bibitem{Asplund:2015eha} 
  C.~T.~Asplund, A.~Bernamonti, F.~Galli and T.~Hartman,
  ``Entanglement Scrambling in 2d Conformal Field Theory,''
  arXiv:1506.03772 [hep-th].
  
   \bibitem{Caputa:2015waa} 
  P.~Caputa, J.~Simon, A.~Stikonas, T.~Takayanagi and K.~Watanabe,
  ``Scrambling time from local perturbations of the eternal BTZ black hole,''
  arXiv:1503.08161 [hep-th].
  %%CITATION = ARXIV:1503.08161;%%
  
  
  \bibitem{NNT}
  M.~Nozaki, T.~Numasawa and T.~Takayanagi,
  ``Holographic Local Quenches and Entanglement Density,''  JHEP {\bf 1305} (2013) 080  [arXiv:1302.5703 [hep-th]]. 

 \bibitem{NCH}
Nielsen, M. A., and I. L. Chuang, 2000, Quantum computation
and quantum information (Cambridge University Press).

\bibitem{DiF}
P. Di Francesco, P. Mathieu, D. Senechal (1997)
"Conformal field theory"

%\cite{Moore:1988ss}
\bibitem{Moore:1988ss} 
  G.~W.~Moore and N.~Seiberg,
  ``Naturality in Conformal Field Theory,''
  Nucl.\ Phys.\ B {\bf 313}, 16 (1989).
  %%CITATION = NUPHA,B313,16;%%
  %187 citations counted in INSPIRE as of 12 Jun 2015
  %\bibitem{Moore:1988uz} 
  G.~W.~Moore and N.~Seiberg,
  ``Polynomial Equations for Rational Conformal Field Theories,''
  Phys.\ Lett.\ B {\bf 212}, 451 (1988).
  %%CITATION = PHLTA,B212,451;%%  

 
  \bibitem{Osborn:2012vt}
  H.~Osborn,
  ``Conformal Blocks for Arbitrary Spins in Two Dimensions,''
  Phys.\ Lett.\ B {\bf 718} (2012) 169
  [arXiv:1205.1941 [hep-th]].
  %%CITATION = ARXIV:1205.1941;%%
  

%\cite{Gaberdiel:1994fs}
\bibitem{Gaberdiel:1994fs} 
  M.~Gaberdiel,
  ``A General transformation formula for conformal fields,''
  Phys.\ Lett.\ B {\bf 325}, 366 (1994)
  [hep-th/9401166].
  %%CITATION = HEP-TH/9401166;%%

 \bibitem{Miyaji:2014mca} 
  M.~Miyaji, S.~Ryu, T.~Takayanagi and X.~Wen,
  ``Boundary States as Holographic Duals of Trivial Spacetimes,''
  JHEP {\bf 1505}, 152 (2015)
  [arXiv:1412.6226 [hep-th]].
  %%CITATION = ARXIV:1412.6226;%%
  
  
  \bibitem{PandoZayas:2014wsa} 
  L.~A.~Pando Zayas and N.~Quiroz,
  ``Left-Right Entanglement Entropy of Boundary States,''
  JHEP {\bf 1501}, 110 (2015)
  [arXiv:1407.7057 [hep-th]].
  %%CITATION = ARXIV:1407.7057;%%
  
    \bibitem{Das:2015oha} 
  D.~Das and S.~Datta,
  ``Universal features of left-right entanglement entropy,''
  arXiv:1504.02475 [hep-th].
  %%CITATION = ARXIV:1504.02475;%%
  
  \bibitem{Brehm:2015lja} 
  E.~M.~Brehm and I.~Brunner,
  ``Entanglement entropy through conformal interfaces in the 2D Ising model,''
  arXiv:1505.02647 [hep-th].


\bibitem{Brustein:1988vb} 
  R.~Brustein, S.~Yankielowicz and J.~B.~Zuber,
  ``Factorization and Selection Rules of Operator Product Algebras in Conformal Field Theories,''
  Nucl.\ Phys.\ B {\bf 313}, 321 (1989).
  %%CITATION = NUPHA,B313,321;%%
  
 \bibitem{Dong:2008ft} 
  S.~Dong, E.~Fradkin, R.~G.~Leigh and S.~Nowling,
  ``Topological Entanglement Entropy in Chern-Simons Theories and Quantum Hall Fluids,''
  JHEP {\bf 0805}, 016 (2008)
  [arXiv:0802.3231 [hep-th]].
  %%CITATION = ARXIV:0802.3231;%% 
  
  \bibitem{Fendley:2006gr} 
  P.~Fendley, M.~P.~A.~Fisher and C.~Nayak,
  ``Topological entanglement entropy from the holographic partition function,''
  J.\ Statist.\ Phys.\  {\bf 126}, 1111 (2007)
  [cond-mat/0609072 [cond-mat.stat-mech]].
  %%CITATION = COND-MAT/0609072;%%

%\cite{Jackson:2014nla}
\bibitem{Jackson:2014nla}
  S.~Jackson, L.~McGough and H.~Verlinde,
  ``Conformal Bootstrap, Universality and Gravitational Scattering,''
  arXiv:1412.5205 [hep-th].
  %%CITATION = ARXIV:1412.5205;%%
  \bibitem{McGough:2013gka} 
  L.~McGough and H.~Verlinde,
  ``Bekenstein-Hawking Entropy as Topological Entanglement Entropy,''
  JHEP {\bf 1311}, 208 (2013)
  [arXiv:1308.2342 [hep-th]].
  %%CITATION = ARXIV:1308.2342;%%
  
  \bibitem{Herwerth:2015pga} 
  B.~Herwerth, G.~Sierra, H.~H.~Tu and A.~E.~B.~Nielsen,
  ``Excited States in Spin Chains from Conformal Blocks,''
  arXiv:1501.07557 [cond-mat.str-el].
  
  \bibitem{Calabrese:2012ew} 
  P.~Calabrese, J.~Cardy and E.~Tonni,
  ``Entanglement negativity in quantum field theory,''
  Phys.\ Rev.\ Lett.\  {\bf 109}, 130502 (2012)
  [arXiv:1206.3092 [cond-mat.stat-mech]].
  %%CITATION = ARXIV:1206.3092;%%
  
 \bibitem{Wen:2015qwa} 
  X.~Wen, P.~Y.~Chang and S.~Ryu,
  ``Entanglement negativity after a local quantum quench in conformal field theories,''
  arXiv:1501.00568 [cond-mat.stat-mech].
  %%CITATION = ARXIV:1501.00568;%% 
  
  \bibitem{Stojevic:2014zta} 
  V.~Stojevic, J.~Haegeman, I.~P.~McCulloch, L.~Tagliacozzo and F.~Verstraete,
  ``Conformal Data from Finite Entanglement Scaling,''
  Phys.\ Rev.\ B {\bf 91}, no. 3, 035120 (2015)
  [arXiv:1401.7654 [quant-ph]].
  %%CITATION = ARXIV:1401.7654;%% 
  
    \bibitem{WIP}
    work in progress
  
    \bibitem{RTorig}
  S.~Ryu and T.~Takayanagi,
  ``Holographic derivation of entanglement entropy from AdS/CFT,''
  Phys.\ Rev.\ Lett.\  {\bf 96} (2006) 181602

\end{thebibliography}
\end{document}